\documentclass[11pt]{article}

\usepackage[final]{acl}

\usepackage{times}
\usepackage{latexsym}

\usepackage[T1]{fontenc}
\usepackage[utf8]{inputenc}

\usepackage{microtype}
\usepackage{adjustbox}
\usepackage{inconsolata}
\usepackage{longtable}
\usepackage{array}
\usepackage{wrapfig}
\usepackage{algorithm}
\usepackage{amssymb}
\usepackage{amsmath}
\usepackage[normalem]{ulem}
\useunder{\uline}{\ul}{}
\usepackage{booktabs}
\usepackage{algpseudocode}
\usepackage{listings}
\usepackage{color}
\usepackage{multirow}
\usepackage{xcolor,tikz}

\usepackage{enumitem}
\usepackage{colortbl}
\usepackage[table,xcdraw]{xcolor}

\definecolor{boldblue}{RGB}{181, 215, 255}
\definecolor{ulblue}{RGB}{218, 235, 255}

\usepackage{graphicx}

\title{Retrieve Only Relevant Tables Whether Few or Many: \\ Adaptive Table Retrieval Method}

\author{
  \textbf{Taehee Kim\textsuperscript{1,2}\thanks{Equal contribution}},
  \textbf{Seungbin Yang\textsuperscript{1}\footnotemark[1]},
  \textbf{Jihwan Kim\textsuperscript{1}},
  \textbf{Jaegul Choo\textsuperscript{1}}
\\
  \textsuperscript{1}KAIST AI, \textsuperscript{2}Letsur \\
  \texttt{\{taeheekim, sby99, jihvvan.kim, jchoo\}@kaist.ac.kr}
}
\begin{document}
\maketitle
\begin{abstract}
Retrieving relevant tables from extensive databases for a given natural language query is essential for accurately answering questions in tasks such as text-to-SQL. 
% Existing table retrieval approaches commonly adopt a top-$k$ retrieval strategy based on query-table similarity. 
Existing table retrieval approaches select a pre-determined set of $k$ tables with the highest similarity to the query.
However, the number of required tables varies across queries and cannot be known in advance. 
% Such a stubborn strategy may either retrieve an undersized set of tables, preventing the model from gathering all that was needed, or retrieve too large a pool of tables, leading to the incorporation of unnecessary ones. 
Enforcing a fixed number of retrieved tables regardless of the query may either retrieve an undersized set, failing to obtain all necessary evidence, or retrieve an oversized pool, including irrelevant tables.
To address this issue, we propose an adaptive table retrieval method that adjusts the number of tables retrieved according to the requirements of each query. 
Specifically, we utilize an adaptive thresholding mechanism to selectively retrieve tables and integrate a sliding-window reranking algorithm to efficiently process a large table corpus. 
Extensive experiments on Spider, BIRD, and Spider 2.0 demonstrate that our method effectively addresses the limitations of the existing retrieval strategy, improving performance in retrieval and downstream tasks.
Our code and data are available at \href{https://github.com/sbY99/Adaptive-Table-Retrieval}{link}.
\end{abstract}

\section{Introduction}

Advances in large language models (LLMs) have led to substantial gains in tasks demanding structured reasoning over tabular data~\citep{gao2023text, xie2024decomposition, yang2024synthesizing}. 
These improvements are critical to real-world applications such as text-to-SQL and open-domain question answering, where leveraging structured data is essential~\citep{zhong2017seq2sql, yu-etal-2018-spider, herzig-etal-2021-open}. 
Retrieval-augmented generation (RAG) approaches address this need, first retrieving tables relevant to a query and then generating an answer conditioned on the retrieved tables~\citep{lewis2020retrieval, pan2022end, kothyari-etal-2023-crush4sql, kang-etal-2024-denoising, kong2024opentab}. 

\begin{figure}[t]
    \centering
        \includegraphics[width=\columnwidth]{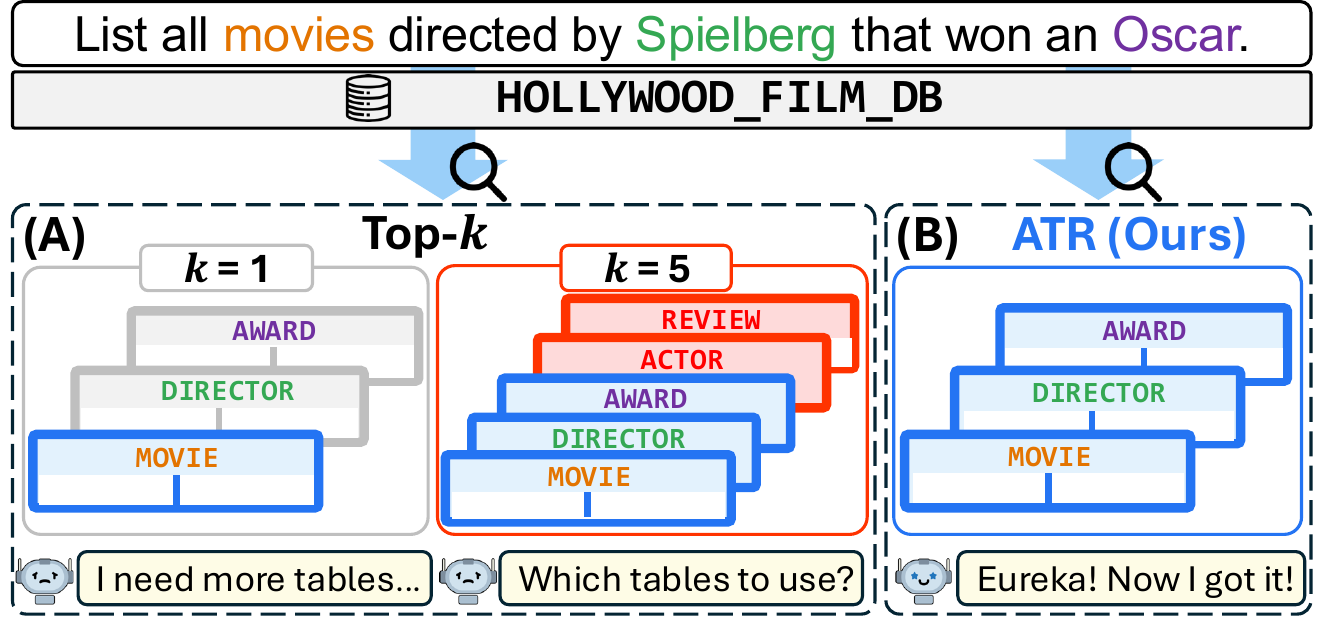}
    \caption{Rather than rely on a rigid fixed $k$ retrieval strategy, ATR retrieves only relevant tables. Gray indicates tables required by the query but not retrieved, red denotes irrelevant tables, and blue highlights retrieved relevant tables.}
    \label{fig:intro}
\end{figure}

% Existing table retrieval methods rely on a top-$k$ retrieval strategy that selects the top-$k$ tables ranked by query-table similarity~\citep{chen-etal-2024-table, wu2025mmqa, zhang-etal-2025-murre}. 
Existing table retrieval methods compute query-table similarity and select a pre-determined $k$ tables with the highest similarity~\citep{chen-etal-2024-table, wu2025mmqa, zhang-etal-2025-murre}. 
% However, this fixed cut-off ignores the fact that the number of tables required by each query is unknown in advance and can vary significantly. 
However, this top-$k$ retrieval strategy overlooks the fact that the number of tables required by each query is unknown in advance and can vary significantly depending on its complexity.
Even within the same database, the ground-truth tables can range from one to several hundred depending on the query.
For example, the problem is evident in Spider 2.0~\citep{lei2025spider}, a realistic enterprise-level text-to-SQL benchmark where the number of ground-truth tables for each query ranges from 1 to 366. 

\begin{figure}[t]
    \centering
    \includegraphics[width=\columnwidth]{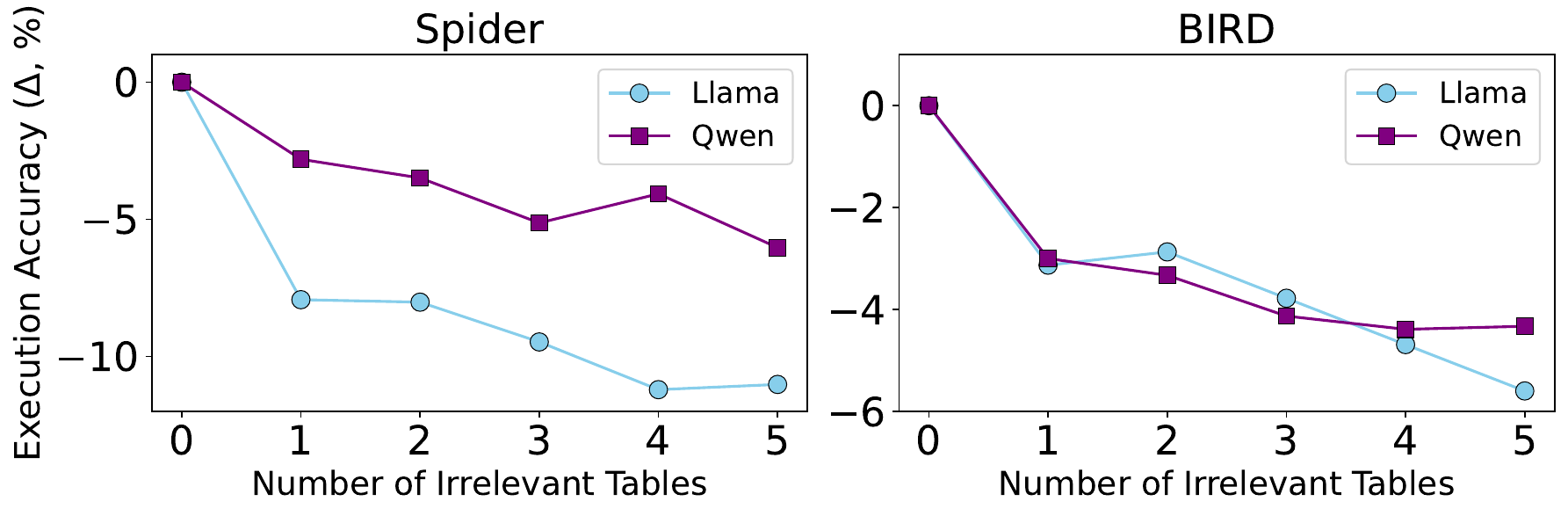}
    \caption{Retrieving irrelevant tables introduces noise, degrading performance on the text-to-SQL task. Execution accuracy consistently decreases as more irrelevant tables are added to the ground-truth tables.}
    % \caption{Retrieving irrelevant tables introduces noise, consistently degrading text-to-SQL performance as irrelevant tables are added to the ground-truth tables.}
    \label{fig:pre}
\end{figure}

Because of the uncertainty, a fixed $k$ retrieval method can miss necessary tables or retrieve too many tables to answer, depending on the value of $k$.
As illustrated in Figure~\ref{fig:intro}-(A), answering the query
\textit{“List all movies directed by Spielberg that won an Oscar.”}
requires three tables: \texttt{MOVIE}, \texttt{DIRECTOR}, and \texttt{AWARD}.
With $k=1$, the retriever misses required tables, whereas $k=5$ inevitably retrieves two irrelevant ones. 
Thus, a small $k$ sacrifices recall and a large $k$ inflates latency and injects noise, degrading downstream performance~\citep{kothyari-etal-2023-crush4sql}. 
Figure~\ref{fig:pre} quantifies how performance decreases as irrelevant tables are added.

To overcome this uncertainty in deciding how many tables a query needs, we introduce \textbf{A}daptive \textbf{T}able \textbf{R}etrieval (ATR). 
As depicted in Figure~\ref{fig:intro}-(B), ATR identifies the exact set of tables, without relying on a fixed $k$. 
ATR uses an adaptive thresholding mechanism to selectively retrieve tables whose logits surpass the query-specific threshold.
A sliding-window reranking algorithm is integrated into ATR to efficiently process large table corpora.
Furthermore, ATR leverages a relevance calibration and semantic grouping loss to effectively model query-table and table-table relevance. 
% As depicted in Figure~\ref{fig:intro}-(B), ATR identifies the exact set of required tables, reranking tables for a given ranked tables. 
% An adaptive threshold mechanism retrieves only those tables with logits exceeding the threshold and a sliding-window algorithm can handle large candidate sets efficiently.
% ATR is guided by a relevance calibration and a semantic grouping loss that jointly model query-table and table-table relevance. 

We evaluate ATR on Spider~\citep{yu-etal-2018-spider}, BIRD~\citep{li2023can}, and Spider 2.0~\citep{lei2025spider}. 
ATR outperforms top-$k$ baselines across all datasets, retrieving fewer irrelevant tables and more essential ones, and improving both retrieval metrics and downstream text-to-SQL execution accuracy.
% The improvements hold whether a query requires a single table or hundreds, demonstrating its robustness.
These improvements hold regardless of the number of tables required, demonstrating robustness.

Our contributions can be summarized as follows:
\begin{itemize}
    \item We demonstrate that fixed $k$ table retrieval overlooks the variation in required tables, resulting in the over-selection of irrelevant tables or the omission of essential ones. 
    % \item We propose ATR, an adaptive table retrieval method that leverages query-dependent thresholding with relevance calibration and semantic grouping losses to precisely retrieve necessary tables, combined with a sliding-window reranking for efficient scaling to large table corpora.
    \item We propose ATR, an adaptive table retrieval method that leverages query-dependent thresholding with relevance calibration and semantic grouping losses to precisely retrieve necessary tables, combined with a sliding-window reranking to ensure scalable and efficient processing across extensive table corpora.
    \item Experiments on Spider, BIRD, and Spider 2.0 show that ATR consistently outperforms strong top-$k$ baselines by precisely retrieving the necessary tables required to answer each query, thereby improving both efficiency and effectiveness in downstream text-to-SQL task.
\end{itemize}

\section{Related Work}

\paragraph{Table Retrieval} Table retrieval is the task of selecting the subset of tables that provide evidence for a natural-language query from databases~\citep{herzig-etal-2021-open, wang2021retrieving, dong2022deepjoin, balaka2025pneuma}. 
\citet{chen-etal-2024-table} introduce methods capturing inter-table relationships and \citet{li2025multifield} propose dynamic weighting of multiple fields based on query.
\citet{li-etal-2025-tailoring} propose a field-aware hybrid matching framework integrating sparse and dense representations.
More recently, studies have explored LLM-based retrieval methods, leveraging LLMs to retrieve relevant tables~\citep{kothyari-etal-2023-crush4sql, zhang-etal-2025-murre, wu2025mmqa}.
% Another line of research is schema linking, which aims to align natural language queries with schema elements~\citep{wang-etal-2020-rat,lei-etal-2020-examining,lee-etal-2025-dcg,wang-etal-2025-linkalign,hao-etal-2025-genlink}.
% Unlike table retrieval which identifies sources from a large-scale collection, schema linking focuses on the fine-grained semantic alignment of query tokens to the internal structure of a database.
Despite these advances, all of the above methods still rely on a fixed $k$ retrieval strategy, failing to adapt to the varying table requirements across queries.

% \citet{kothyari-etal-2023-crush4sql} and \citet{zhang-etal-2025-murre} propose table retrieval methods that rewrite user queries to enhance table retrieval accuracy. 
% \citet{chen-etal-2024-table} and \citet{wu2025mmqa} introduce methods that capture inter-table relationships, enhancing the coherence of retrieved tables.
% \citet{li2025multifield} propose a retrieval method that dynamically weights multiple fields of semi-structured data according to query semantics.
% Despite these advances, all of the above still rely on a top-$k$ cut-off: with a small $k$ they may omit essential tables, whereas a large $k$ retrieves many irrelevant ones, degrading downstream accuracy and efficiency. 

\paragraph{Adaptive Retrieval Strategy}
Recent RAG research has shown growing interest in adaptive retrieval methods, which adjust the size of text chunks based on query~\citep{jiang-etal-2023-active, asai2023self, jeong2024adaptive}.
These approaches assess query complexity and selectively increase the retrieval budget accordingly.
However, such methods are designed exclusively for text and do not consider structured tabular data.
Moreover, they involve iterative interactions with an LLM, increasing inference cost.
To overcome these limitations, ATR adopts an adaptive thresholding to retrieve a query-dependent number of tables without iterative generator interactions and explicitly learns table representations optimized for tabular data.
% To address this issue, ATR applies an adaptive table retrieval method to rerank a list of tables, allowing it to retrieve a query-dependent number of tables, in contrast to other approaches that depend on generator interactions.

\paragraph{Text-to-SQL}
Text-to-SQL is the task of generating SQL queries from natural language questions, enabling effective access to structured databases~\citep{zhong2017seq2sql, yu-etal-2018-spider, hong-etal-2024-knowledge}.
Recent studies in text-to-SQL adopt retrieval-augmented approaches that integrate table retrieval with text-to-SQL generation to handle large-scale database scenarios~\citep{kothyari-etal-2023-crush4sql, kong2024opentab, chen-etal-2024-table}.
Additionally, large-scale datasets recently introduced by \citet{chen2024beaver} and  \citet{lei2025spider} focus on enterprise-level text-to-SQL tasks.
In line with these developments, our work contributes an adaptive retrieval framework that efficiently scales to large table corpora, significantly enhancing text-to-SQL performance and efficiency in resource-intensive contexts.

\begin{figure*}[t]
    \centering
    \includegraphics[width=\textwidth]{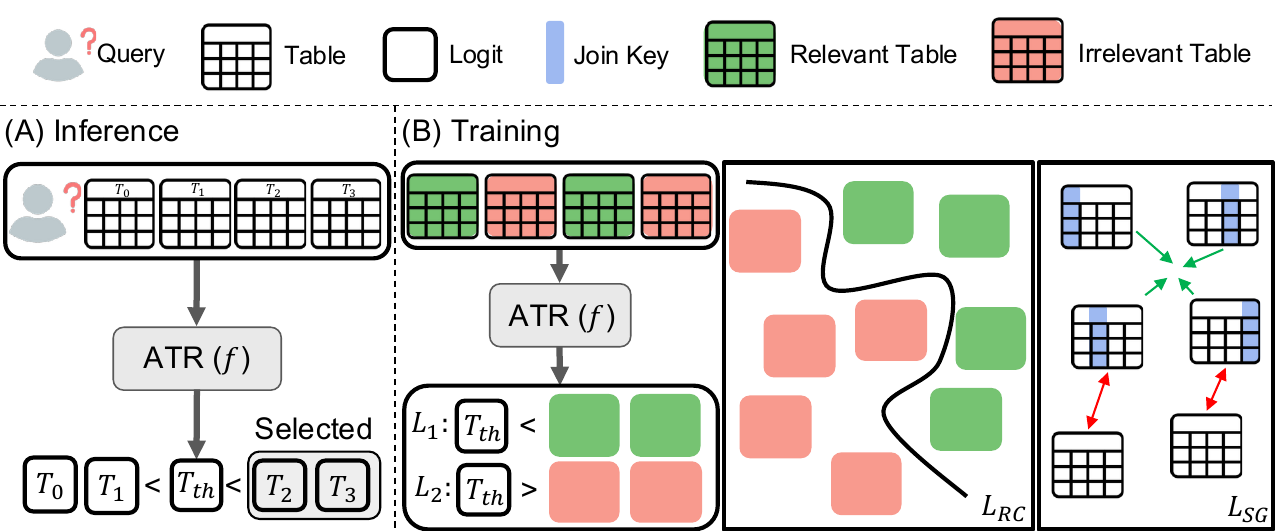}
    \caption{Overview of the ATR framework. (A) Inference: ATR takes a query and candidate tables as input to compute relevance logits. It dynamically selects the subset of tables (e.g., $T_2, T_3$) whose logits exceed that of the adaptive threshold token ($T_{th}$), thereby automatically determining the specific number of tables needed for the query. (B) Training: ATR is optimized using three complementary objectives: 1) Adaptive Thresholding ($L_{AT}$) trains the threshold token to serve as a decision boundary, enforcing relevant tables to score higher ($L_1$) and irrelevant ones lower ($L_2$) than $T_{th}$; 2) Relevance Calibration ($L_{RC}$) maximizes the logit margin between relevant and irrelevant tables; and 3) Semantic Grouping ($L_{SG}$) adopts contrastive learning to pull embeddings of joinable tables that are sharing join keys closer together while pushing non-joinable ones apart.}
    \label{fig:main}
\end{figure*}

\section{Problem Definition}
% We formulate the table retrieval task as follows.
We formally define the table retrieval task as follows.
% Given a query $q$ and a table corpus $\mathcal{C} = \{t_i\}_{i=1}^{N}$, where each table $t_i$ contains structured information, the objective of the table retrieval task is to find a subset of tables $\mathcal{C}_q \subseteq \mathcal{C}$ that collectively satisfies the informational need expressed by the query $q$.
Given a query $q$ and a table corpus $\mathcal{C} = \{t_i\}_{i=1}^{N}$, where each table $t_i$ contains structured information, the objective of the table retrieval task is to find a subset of tables $\mathcal{C}_q \subseteq \mathcal{C}$ that cover the information required to accurately answer the query.

A retrieval function $f$ ranks tables in corpus $\mathcal{C}$ by descending order of the relevance scores $s(q,t_i)$, computed based on the query-table relevance.
The top-$k$ tables $\hat{\mathcal{C}}_q$ are selected according to $s(q, t_i)$.

\begin{equation*}
\scalebox{1.0}{$
    \begin{gathered}
        \hat{\mathcal{C}}_q = \{\,t_q^{(i)}\}_{i=1}^{k} = f\bigl(q, \mathcal{C}\bigr),
        \\[3pt]
        \quad\text{where}\quad
        s\bigl(q, t_q^{(n)}\bigr) \;>\; s\bigl(q, t_q^{(m)}\bigr)
        \quad \forall\, n < m
    \end{gathered}
$}
\end{equation*}

\section{Adaptive Table Retrieval}
\label{ATR:main}
% ATR adaptively selects the number of tables that each query needs.
% Whereas standard top-$k$ retrieval always returns a fixed $k$ tables regardless of the query, ATR infers a query-specific number of tables $k_q$ to retrieve:
While standard top-$k$ retrieval always returns a fixed $k$ of tables, ATR adaptively selects a query-specific number of tables $k_q$ required for each query:
\begin{equation*}
    \hat{\mathcal{C}}_q
    \;=\;
    \{\,t_q^{(i)}\}_{i=1}^{k_q}
    \;=\;
    \mathrm{ATR}\bigl(q,\mathcal{C}\bigr)
\end{equation*}

% ATR—a transformer encoder—uses the query, the candidate tables, and two additional embedding tokens as input. 
% We use the hidden states from each table and special tokens to infer $\hat{\mathcal{C}}_q$. 
% $k_q$ is decided by a comparison of logits from the hidden states between each table and the special token. 
ATR is a transformer encoder that takes as input a query along with tables. 
It encodes tables into hidden representations conditioned on query relevance. 
The final retrieved set $\hat{\mathcal{C}}_q$ and the number of retrieved tables $k_q$ are adaptively determined by comparing the logit of each table's hidden states with a query-dependent adaptive threshold. 
To effectively capture both query-table and inter-table relevance, we use two complementary objectives: the \emph{relevance calibration} loss that sharpens query-table alignment, and the \emph{semantic grouping} loss that pulls the embeddings of joinable tables closer. 
For efficient inference over large table corpora, we propose a sliding-window reranking method that refines the ranking without exhaustively scoring every table at once.
% Figure~\ref{fig:main} provides an overview of the ATR architecture and its training strategies.
Figure~\ref{fig:main} illustrates the operational mechanism of ATR, detailing the adaptive retrieval process and its training objectives.

\paragraph{Adaptive Thresholding}
\label{method:adaptive_thresholding}
ATR is trained to separate relevant tables from irrelevant tables for each query through an adaptive thresholding mechanism inspired by~\citet{zhou2021document}.
% We prepend a \emph{threshold token} $T_{th}$ to the input sequence, followed by the natural language query and tables. 
% Every table begins with a \emph{table token} $T_{tab}$, which serves as a representative embedding for the table.
% Each table contains metadata, including its database name, table name, and column names.
We prepend a \emph{threshold token} $T_{th}$ to the input sequence to represent the adaptive boundary between relevant and irrelevant tables.
For each table $t_i$ consisting of the database name, table name, and column names, we prepend a \emph{table token} $T_i$ to serve as its representative embedding.
The resulting input sequence is structured as: 
$[T_{th}; q; T_{1}; t_1; \dots; T_{n}; t_n]$.

ATR computes logits $logit_{T_{th}}$ and $logit_{T_{i}}$ from the hidden states of special tokens $T_{th}$ and $T_{i}$, respectively.
While training, we enforce that $logit_{T_{i}}$ is bigger than $logit_{T_{th}}$ when the table is relevant, and is smaller than $logit_{T_{th}}$ otherwise. 
The loss for adaptive thresholding is as follows:

\begin{equation*}
    \begin{gathered}
    L_1 = -\sum_{r \in \mathcal{T}^{+}} \log \frac{\exp(\text{logit}_r)}{\sum_{r' \in \mathcal{T}^{+} \cup \{{T_{th}}\}} \exp(\text{logit}_{r'})} \\[4pt]
    L_2 = -\log \frac{\exp(\text{logit}_{T_{th}})}{\sum_{r' \in \mathcal{T}^{-} \cup \{{T_{th}}\}} \exp(\text{logit}_{r'})} \\[4pt]
    L_{AT} =  \alpha L_1 + \beta L_2
    \end{gathered}
\end{equation*}
where $\mathcal{T}^{+}$ denotes the set of relevant table tokens, and $\mathcal{T}^{-}$ denotes the set of irrelevant table tokens.

$L_1$ raises logits of query-relevant tables above the threshold logit $logit_{T_{th}}$, creating a clear margin from irrelevant tables. 
Since a query can have multiple relevant tables, we compute a binary cross-entropy loss for each relevant table and sum the results. 
In contrast, $L_2$ suppresses the logits of irrelevant tables below $logit_{T_{th}}$ by treating the threshold token as the correct class. 
The threshold logit thus becomes a query-dependent decision boundary that distinguishes between relevant and irrelevant tables. 
The hyper-parameters $\alpha$ and $\beta$ weight the relative contributions of the two loss terms. 

\paragraph{Relevance Calibration}
Adaptive thresholding assigns a query-specific cut-off by computing a logit for every table and comparing it with the logit of the threshold token.
To sharpen the distinction between relevant tables $\mathcal{T}^{+}$ and irrelevant tables $\mathcal{T}^{-}$, we maximize the logit gap between $\mathcal{T}^{+}$ and $\mathcal{T}^{-}$ using a binary cross-entropy (BCE) loss:
\begin{equation*}
\scalebox{0.9}{$
\begin{split}
L_{RC} = -\frac{1}{|\mathcal{T}^{+} \cup \mathcal{T}^{-}|} &\left(\sum_{r \in \mathcal{T}^{+}}\log\left(\sigma(\text{logit}_r)\right)\right.\\
&\quad\left.+ \sum_{r \in \mathcal{T}^{-}}\log\left(1 - \sigma(\text{logit}_r)\right)\right)
\end{split}$}
\end{equation*}
where $\sigma$ denotes the sigmoid function. 

This relevance calibration loss aligns each query with its relevant tables, giving ATR a signal to distinguish them from irrelevant ones and improving its discriminative capability.

\begin{table*}[t]
\centering
\resizebox{\textwidth}{!}{%
\begin{tabular}{lccccccccccccc}
\toprule
                    &            & \multicolumn{4}{c}{\textbf{Spider}}                           & \multicolumn{4}{c}{\textbf{BIRD}}                             & \multicolumn{4}{c}{\textbf{Spider 2.0}}                       \\ \cline{2-14} 
                    & \textit{k} & \textbf{P}    & \textbf{R}    & \textbf{CR}   & \textbf{F1}   & \textbf{P}    & \textbf{R}    & \textbf{CR}   & \textbf{F1}   & \textbf{P}    & \textbf{R}    & \textbf{CR}   & \textbf{F1}   \\ \hline
\multicolumn{14}{l}{\textit{\textbf{Encoder-based}}}                                                                                                                                                                             \\ \hline
Contriever          & 3          & 46.5          & 94.0          & 89.2          & 60.1          & 45.6          & 72.9          & 55.1          & 54.6          & 24.7          & 37.6          & 24.8          & 27.4          \\
                    & 5          & 29.3          & 97.4          & 95.5          & 43.7          & 31.2          & 82.1          & 68.6          & 44.2          & 19.1          & 45.1          & 32.6          & 24.4          \\
                    & 10         & 15.0          & 99.2          & 98.7          & 25.6          & 18.5          & 96.1          & 92.6          & 30.5          & 13.3          & 56.5          & 44.8          & 19.5          \\
UAE                 & 3          & 46.1          & 93.0          & 88.0          & 59.5          & 48.2          & 79.0          & 63.5          & 58.3          & 27.1          & 40.6          & 27.6          & 29.7          \\
                    & 5          & 29.5          & 97.8          & 96.4          & 44.1          & 32.7          & 87.2          & 77.3          & 46.5          & 20.7          & 48.5          & 34.9          & 26.3          \\
                    & 10         & 15.0          & 99.4          & 99.0          & 25.6          & 18.7          & 97.6          & 95.1          & 30.9          & 14.2          & 60.7          & 49.0          & 20.7          \\ \hline
\multicolumn{14}{l}{\textit{\textbf{LLM-based}}}                                                                                                                                                                                 \\ \hline
RankZephyr          & 3          & 47.7          & 95.9          & 92.8          & 61.5          & 47.9          & 76.6          & 60.4          & 57.3          & 26.0          & 39.3          & 26.4          & 28.7          \\
                    & 5          & 29.5          & 98.2          & 96.9          & 44.1          & 32.1          & 84.4          & 72.4          & 45.5          & 19.7          & 46.7          & 33.8          & 25.2          \\
                    & 10         & 15.1          & \underline{99.5}    & \underline{99.2}    & 25.7          & 18.5          & 96.5          & 93.1          & 30.6          & 13.6          & 57.3          & 45.5          & 19.8          \\
Murre     & 3          & 40.5          & 82.9          & 73.8          & 52.5          & 44.6          & 72.5          & 54.6          & 53.7          & 28.1          & 42.3          & 29.4          & \underline{31.0}    \\
                    & 5          & 26.9          & 90.3          & 85.5          & 40.3          & 30.1          & 80.5          & 66.4          & 42.8          & 21.8          & 51.1          & 36.8          & 28.0          \\
                    & 10         & 14.6          & 96.8          & 95.1          & 24.8          & 16.7          & 88.2          & 78.3          & 27.6          & 14.8          & 61.9          & 48.5          & 21.5          \\ \hline
\multicolumn{14}{l}{\textit{\textbf{Reranker-based}}}                                                                                                                                                                            \\ \hline
% BGE-M3 + Contriever & 3          & 46.0          & 93.1          & 87.7          & 59.5          & 39.4          & 64.7          & 44.4          & 47.6          & 24.7          & 33.4          & 21.2          & 24.7          \\
%                     & 5          & 28.9          & 96.7          & 93.7          & 43.3          & 29.5          & 78.9          & 64.2          & 41.9          & 19.2          & 40.2          & 26.7          & 22.1          \\
%                     & 10         & 14.9          & 99.1          & 98.1          & 25.4          & 17.4          & 91.3          & 83.6          & 28.8          & 14.2          & 52.6          & 40.5          & 18.3          \\
% BGE-M3 + UAE        & 3          & 45.7          & 92.5          & 87.1          & 59.1          & 37.4          & 61.2          & 40.7          & 45.2          & 24.2          & 33.9          & 21.6          & 24.9          \\
%                     & 5          & 28.8          & 96.5          & 93.2          & 43.2          & 28.3          & 76.1          & 61.3          & 40.4          & 19.5          & 43.4          & 29.4          & 23.3          \\
%                     & 10         & 14.9          & 98.8          & 97.6          & 25.4          & 17.2          & 90.5          & 82.2          & 28.4          & 14.0          & 54.1          & 40.7          & 18.5          \\
JAR (w. Contriever)    & 3          & 48.1          & 96.5          & 93.6          & 62.0          & \textbf{54.4} & 87.4          & 76.3          & 65.0          & \textbf{29.4} & 42.3          & 26.9          & \textbf{31.6} \\
                    & 5          & 29.7          & 98.5          & 97.2          & 44.3          & 35.1          & 92.5          & 85.9          & 49.7          & 21.9          & 48.6          & 34.0          & 27.0          \\
                    & 10         & 15.1          & \underline{99.5}    & \underline{99.2}    & 25.7          & 18.8          & 97.6          & \underline{96.0}    & 31.0          & 14.0          & 55.1          & 41.4          & 19.5          \\
JAR (w. UAE)           & 3          & 48.4          & 96.5          & 94.1          & 62.3          & 53.6          & 86.3          & 74.8          & 64.4          & \underline{29.1}    & 41.6          & 27.1          & 30.8          \\
                    & 5          & 29.9          & 99.1          & 98.5          & 44.7          & 34.4          & 91.1          & 82.9          & 48.9          & 21.8          & 47.9          & 33.8          & 26.5          \\
                    & 10         & 15.1          & \underline{99.5}    & \underline{99.2}    & 25.7          & 18.7          & 97.2          & 95.2          & 30.9          & 14.3          & 56.4          & 43.9          & 19.7          \\ \bottomrule
\multicolumn{14}{l}{\textit{\textbf{Ours}}}                                                                                                                                                                                      \\ \hline
ATR (w. Contriever)    &            & \textbf{69.6} & \underline{99.5}    & \underline{99.2}    & \textbf{78.3} & \underline{54.0}    & \underline{98.2}    & \underline{96.0}    & \textbf{65.8} & 21.9          & \underline{72.4}    & \underline{64.4}    & 27.8          \\
ATR (w. UAE)           &            & \underline{69.3}    & \textbf{99.6} & \textbf{99.4} & \underline{78.1}    & 52.8          & \textbf{98.6} & \textbf{97.1} & \underline{65.1}    & 19.9          & \textbf{75.4} & \textbf{68.7} & 26.7          \\ \bottomrule
\end{tabular}%
}
\caption{Retrieval performance comparison with baseline methods, evaluated using Precision (P), Recall (R), Complete Recall (CR), and F1 scores (F1). 
We use SGPT-5.8B as the backbone model for Murre.
ATR consistently outperforms all baselines across datasets. The best and second-best scores for each metric are highlighted in \textbf{bold} and \underline{underlined}, respectively.}
\label{Experiments:Main Result}
\end{table*}

\paragraph{Semantic Grouping}
Relationships between tables, especially whether tables can be joined, are critical for accurate multi-table retrieval~\citep{chen-etal-2024-table, wu2025mmqa}.
To consider these dependencies between tables, we add a contrastive learning objective~\citep{1640964, chen2020simple} that pulls embeddings of \emph{joinable} tables closer together while separating embeddings of \emph{non-joinable} tables by a fixed margin.
% Let $e_i$ be the embedding of $t_{i}$ table and let $g_i$ denote its joinability group; the semantic grouping loss $L_{SG}$ is defined as follows:
Let $e_i$ be the embedding of table $t_{i}$ and let $g_i$ denote its label, where tables sharing the same label are joinable; the semantic grouping loss $L_{SG}$ is defined as:

\begin{equation*}
\begin{split}
L_{SG} &= \frac{1}{|\mathcal{P}|}\sum_{(i,j) \in \mathcal{P}} 
\Bigl[\,\mathbb{I}(g_i = g_j)\,\|e_i - e_j\|_2^2 \\
&\quad + \mathbb{I}(g_i \neq g_j)\,\max\bigl(0,\, m - \|e_i - e_j\|_2\bigr)^2\Bigr]
\end{split}
\end{equation*}

where $\mathcal{P}$ represents all unique pairs of $\mathcal{C}$, and $m$ denotes the margin hyper-parameter. 
$L_{SG}$ encourages the embeddings to reflect relational connectivity and, in turn, promotes the retrieval of semantically coherent table sets. 

% Finally, the ATR objective function can be defined as follows:
Finally, the ATR objective function is defined as follows:
\begin{equation*}
L_{ATR} = L_{AT} + \lambda L_{RC} + \gamma L_{SG}
\end{equation*}
where $\lambda$ and $\gamma$ are hyper-parameters that adjust the magnitude of the losses.
These losses enable ATR to adaptively retrieve tables, considering the relevance between the query and the tables, as well as the relevance between tables.

\paragraph{Sliding Window Reranking}
Since the encoder used in ATR has a strict length constraint and a quadratic complexity with respect to input length, directly processing large numbers of tables at once is computationally impractical.
To mitigate this inefficiency, ATR uses a \emph{sliding window reranking} strategy. 
Given a window size $W$ and a retention size $R$ with $R <$  $W$, ATR processes the tables $\mathcal{C}$ from lowest to highest in their initial ranking. 
First, in $W$ lowest-ranked tables, ATR computes logits for every token $T_{i}$ and for the threshold token $T_{th}$, and keeps the top $R$ tables by logit value. 
Then the retained set is merged with the next $W-R$ tables in the original order.
If the threshold logit $logit_{T_{th}}$ ranks lower than $R$ within $W$, its rank is finalized. 
This iterative process continues until all candidate tables have been processed.
Eventually, all the tables that outrank the threshold are included in the final table list. 
Since ATR reranks subsets of tables within overlapping windows, the method avoids the cost of reranking the full list at once. 
Pseudo-code for this sliding window reranking is presented in Algorithm \ref{alg:rerank} and an illustrative example is provided in Figure~\ref{fig:sliding_window_example}. 

\begin{table*}[t]
\centering
\resizebox{\textwidth}{!}{%
\begin{tabular}{lcccccccccc}
\toprule
 &  & \multicolumn{3}{c}{\textbf{Spider}} & \multicolumn{3}{c}{\textbf{BIRD}} & \multicolumn{3}{c}{\textbf{Spider 2.0}} \\ \cline{3-11} 
 & \textit{k} & \textbf{\begin{tabular}[c]{@{}c@{}}Llama\\ (8B/70B)\end{tabular}} & \textbf{\begin{tabular}[c]{@{}c@{}}Qwen\\ (7B/32B)\end{tabular}} & \textbf{\begin{tabular}[c]{@{}c@{}}Gemma\\ (4B/27B)\end{tabular}} & \textbf{\begin{tabular}[c]{@{}c@{}}Llama\\ (8B/70B)\end{tabular}} & \textbf{\begin{tabular}[c]{@{}c@{}}Qwen\\ (7B/32B)\end{tabular}} & \textbf{\begin{tabular}[c]{@{}c@{}}Gemma\\ (4B/27B)\end{tabular}} & \textbf{\begin{tabular}[c]{@{}c@{}}Llama\\ (8B/70B)\end{tabular}} & \textbf{\begin{tabular}[c]{@{}c@{}}Qwen\\ (7B/32B)\end{tabular}} & \textbf{\begin{tabular}[c]{@{}c@{}}Gemma\\ (4B/27B)\end{tabular}} \\ \hline
\multicolumn{11}{l}{\textit{\textbf{Encoder-based}}} \\ \hline
Contriever & 3 & 54.6 / 64.8 & 65.4 / 66.3 & 58.0 / 67.4 & 21.7 / 40.4 & 35.4 / 46.0 & 21.2 / 38.9 & \underline{1.1} / 3.4 & 0.5 / 3.9 & 0.2 / 2.5 \\
 & 5 & 53.3 / 65.0 & 65.7 / 68.6 & 58.3 / 67.7 & 22.4 / 44.5 & 37.0 / 49.8 & 24.1 / 39.7 & \underline{1.1} / 3.2 & 1.4 / 3.7 & 0.7 / 3.2 \\
 & 10 & 54.6 / 67.1 & 66.0 / 70.1 & \underline{60.3} / 67.6 & 24.3 / 47.4 & 39.1 / 53.0 & 25.6 / 43.2 & \underline{1.1} / 3.5 & 0.9 / 4.1 & 0.2 / 3.4 \\
UAE & 3 & 55.7 / 65.0 & 63.6 / 68.0 & 52.5 / 68.1 & 22.6 / 40.7 & 35.1 / 45.4 & 20.1 / 38.8 & 0.7 / 3.5 & 1.1 / 3.4 & 0.5 / 3.4 \\
 & 5 & 54.9 / 66.1 & 66.0 / 69.6 & 52.8 / 70.2 & 23.9 / 45.0 & 37.8 / 49.5 & 22.0 / 43.5 & 0.9 / 3.9 & 1.1 / 3.4 & \underline{0.9} / 4.1 \\
 & 10 & 54.7 / 67.3 & 66.1 / 70.6 & 54.1 / 69.2 & 26.0 / 47.9 & 40.4 / 52.2 & 22.2 / 46.8 & 1.4 / 4.1 & 1.4 / 3.4 & 0.5 / 3.9 \\ \hline
\multicolumn{11}{l}{\textit{\textbf{LLM-based}}} \\ \hline
RankZephyr & 3 & 53.3 / 64.8 & 63.1 / 66.6 & 54.2 / 69.4 & 22.4 / 40.7 & 34.5 / 45.8 & 21.7 / 43.2 & \underline{1.1} / 3.2 & 1.4 / 4.1 & 0.2 / 2.5 \\
 & 5 & 54.8 / 65.9 & 65.4 / 67.2 & 58.2 / 70.2 & 23.0 / 45.0 & 38.1 / 46.2 & 23.2 / 45.5 & 0.9 / 3.0 & 1.1 / 4.1 & 0.7 / 3.0 \\
 & 10 & 57.5 / 67.0 & 66.4 / 67.3 & 59.0 / 70.7 & 25.4 / 47.5 & 39.3 / 49.5 & 24.0 / 46.2 & \underline{1.1} / 3.4 & 1.4 / 3.4 & 0.5 / 4.4 \\
Murre & 3 & 51.2 / 65.3 & 63.4 / 64.6 & 53.6 / 69.3 & 21.8 / 44.5 & 34.2 / 44.2 & 20.9 / 42.8 & 0.9 / 4.1 & 1.4 / 3.2 & \underline{0.9} / 4.1 \\
 & 5 & 53.7 / 65.6 & 64.1 / 65.8 & 54.1 / 70.5 & 22.6 / 44.8 & 37.5 / 45.4 & 20.7 / 44.5 & 0.9 / 3.0 & 1.4 / 3.7 & \underline{0.9} / 3.0 \\
 & 10 & 54.5 / 65.9 & 65.3 / 68.0 & 55.0 / 70.2 & 25.3 / 46.4 & 39.4 / 48.4 & 21.2 / \underline{47.0} & 0.7 / 4.1 & 1.4 / 4.4 & 0.7 / \underline{4.6} \\ \hline
\multicolumn{11}{l}{\textit{\textbf{Reranker-based}}} \\ \hline
JAR (w. Contriever) & 3 & 53.9 / 65.9 & 66.6 / 68.8 & 54.8 / 69.1 & 25.4 / 45.3 & 40.1 / 49.7 & 23.9 / 44.4 & 0.7 / 3.7 & 0.9 / 4.1 & 0.5 / 3.0 \\
 & 5 & 54.8 / 66.6 & 67.5 / 69.7 & 59.4 / 67.8 & 26.8 / 46.4 & 43.4 / 52.0 & \underline{27.4} / 43.2 & \underline{1.1} / 3.9 & 0.7 / 3.7 & \underline{0.9} / 3.2 \\
 & 10 & 56.0 / 66.8 & 66.6 / 69.2 & 59.5 / 67.7 & 26.7 / 47.5 & 43.0 / 53.0 & 24.8 / 44.9 & 0.9 / 3.0 & 1.4 / 4.4 & 0.5 / 4.1 \\
JAR (w. UAE)  & 3 & 54.6 / 67.2 & 66.7 / 66.8 & 55.1 / 69.2 & 24.8 / 44.7 & 38.7 / 44.9 & 22.6 / 44.3 & 0.9 / 3.9 & \underline{1.6} / 4.4 & 0.5 / 2.8 \\
 & 5 & 56.0 / 67.4 & 67.0 / 69.7 & 55.3 / 70.2 & 25.3 / 45.4 & 39.2 / 48.4 & 23.1 / 44.7 & 0.9 / 4.1 & 1.4 / 4.1 & \underline{0.9} / 4.4 \\
 & 10 & 55.6 / 66.9 & 64.6 / 67.7 & 56.6 / \underline{70.8} & 26.6 / 46.9 & 41.3 / 52.2 & 24.2 / 46.6 & 0.9 / 3.0 & \underline{1.6} / 3.7 & \underline{0.9} / 3.2 \\ \hline
\multicolumn{11}{l}{\textit{\textbf{Ours}}} \\ \hline
ATR (w. Contriever) &  & \underline{58.7} / \underline{67.8} & \underline{69.7} / \textbf{71.5} & \textbf{60.7} / 67.8 & \underline{28.6} / \textbf{49.9} & \textbf{45.0} / \underline{53.3} & \textbf{28.8} / 45.1 & \underline{1.1} / \underline{4.4} & 1.4 / \textbf{5.7} & \textbf{1.1} / \textbf{4.8} \\
ATR (w. UAE) &  & \textbf{59.8} / \textbf{68.2} & \textbf{69.8} / \underline{71.4} & 59.0 / \textbf{71.1} & \textbf{29.0} / \underline{48.7} & \underline{43.8} / \textbf{53.9} & 25.6 / \textbf{49.7} & \textbf{1.6} / \textbf{4.5} & \textbf{2.1} / \underline{4.9} & \textbf{1.1} / \underline{4.6} \\ \hline
\multicolumn{1}{l}{\textit{\textbf{Oracle}}} &  & 66.6 / 70.8 & 75.6 / 75.2 & 66.5 / 71.7 & 31.8 / 53.5 & 50.6 / 58.0 & 32.7 / 50.9 & 4.4 / 7.2 & 3.5 / 7.4 & 1.4 / 7.6 \\ \bottomrule
\end{tabular}%
}
\caption{Text-to-SQL execution accuracy comparison across different table retrieval methods. ATR consistently outperforms baseline retrievers on the Spider, BIRD, and Spider 2.0 datasets.}
\label{tab:end-to-end_performance}
\end{table*}

\section{Experiments}
% In this section, we introduce the datasets, metrics, models, and baselines used to evaluate the performance of ATR.
% The experiments aim to validate the effectiveness of ATR that overcomes the inherent limitations of existing table retrieval strategies and to quantify the impact of improved retrieval performance on the downstream task.

\subsection{Setup}

\paragraph{Dataset}
We use three datasets: Spider~\citep{yu-etal-2018-spider}, BIRD~\citep{li2023can}, and Spider 2.0~\citep{lei2025spider}.
Spider and BIRD are widely used benchmarks for text-to-SQL.
We adopt the "union" setting, merging all databases into a single corpus~\citep{kothyari-etal-2023-crush4sql, zhang-etal-2025-murre}.
Spider 2.0 is a benchmark composed of enterprise text-to-SQL workflows derived from a large-scale database.
We use Spider 2.0-Lite\footnote{For simplicity, we refer to Spider 2.0-Lite as Spider 2.0.}, a subset featuring multiple SQL dialects, and apply the union setting grouped by dialect.
We note that ATR training utilizes only the training sets of Spider and BIRD.
Spider and BIRD contain queries requiring between 1 and 5 tables, while Spider 2.0 queries range from 1 to 366 tables.
Detailed data pre-processing methods are described in Appendix~\ref{app:dataset_processing}, and dataset statistics are presented in Table~\ref{tab:datasets}.

\paragraph{Task and Metrics}
We evaluate ATR on two tasks: table retrieval and text-to-SQL generation.
In table retrieval, given a natural language query $q$, a model retrieves a set of tables $\hat{\mathcal{C}_q}\subset\mathcal{C}$ from the table corpus $\mathcal{C}$. 
We report \textit{precision}, \textit{recall}, and \textit{F1}, as well as \textit{complete recall} by comparing $\hat{\mathcal{C}_q}$ with the ground-truth set $\mathcal{C}_q$, following \citet{zhang-etal-2025-murre}.
% Retrieving all the ground-truth tables is critical for the table retrieval task, we report \textit{recall} and \textit{complete recall} by comparing $\hat{\mathcal{C}_q}$ with the ground-truth set $\mathcal{C}_q$ following \citet{zhang-etal-2025-murre}.
In the text-to-SQL task, the query $q$ and its retrieved tables $\hat{\mathcal{C}_q}$ are fed to a generator that produces a SQL statement.
We measure \emph{execution accuracy}: the proportion of generated SQL queries whose execution results match those of the reference SQL~\citep{yu-etal-2018-spider}.
To evaluate downstream performance, we ensure that the only difference between retrieval methods is the input tables, allowing a precise assessment of how retrieval performance influences the downstream results.

\paragraph{Models}
We use ModernBERT-large~\citep{modernbert}, a bidirectional encoder-only transformer, as the backbone model for ATR.
For table embedding models, we use Contriever~\citep{izacard2021unsupervised} and UAE-Large-V1~\citep{li-li-2024-aoe}.
For LLM-based retrieval, we leverage SGPT-5.8B~\citep{muennighoff2022sgpt} and RankZephyr-7B-V1~\citep{pradeep2023rankzephyreffectiverobustzeroshot}.
% For table embedding models, we use Contriever~\citep{izacard2021unsupervised}\footnote{\scriptsize \url{https://huggingface.co/facebook/contriever-msmarco}} and UAE-Large-V1~\citep{li-li-2024-aoe}.\footnote{\scriptsize\url{https://huggingface.co/WhereIsAI/UAE-Large-V1}}
% For LLM-based retrieval, we leverage RankZephyr~\citep{pradeep2023rankzephyreffectiverobustzeroshot}\footnote{\scriptsize\url{https://huggingface.co/castorini/rank_zephyr_7b_v1_full}} and SGPT~\citep{muennighoff2022sgpt}\footnote{\scriptsize\url{https://huggingface.co/Muennighoff/SGPT-5.8B-weightedmean-msmarco-specb-bitfit}}.
For SQL generation, we utilize Llama-3.1-8B/70B-Instruct~\citep{grattafiori2024llama}, Qwen2.5-Coder-7B/32B-Instruct~\citep{hui2024qwen2}, and Gemma-3-4B/27B-IT~\citep{team2025gemma} as generators. 
Results for additional models, including proprietary models, are demonstrated in Appendix~\ref{app:additional_retrieval_results} and ~\ref{app:additional_end-to-end_results}.
% Appendix~\ref{app:additional_retrieval_results} and ~\ref{app:additional_end-to-end_results} demonstrate additional model results, including proprietary models.

\begin{figure*}[ht]
    \centering
\includegraphics[width=\textwidth,height=4cm]{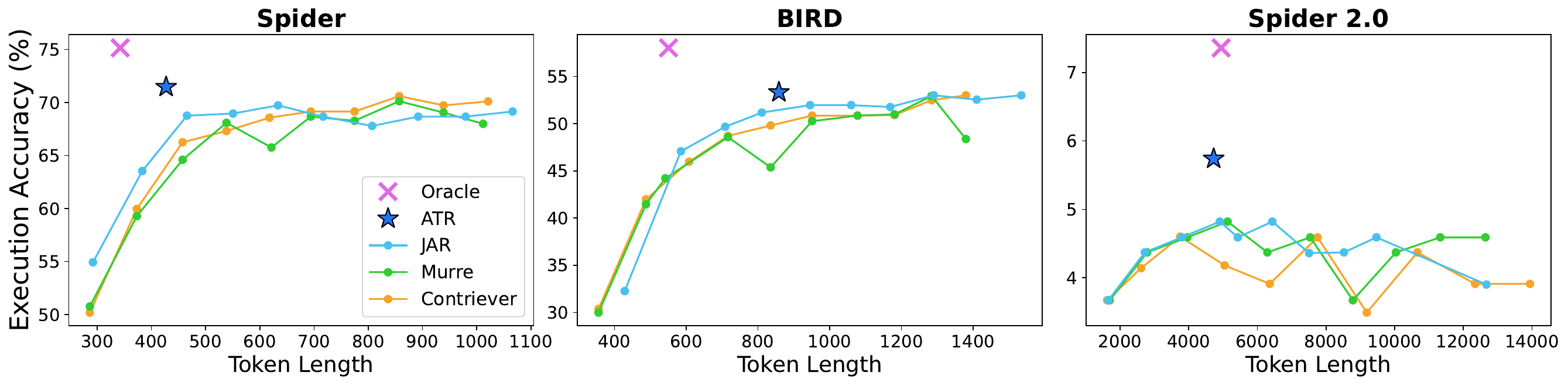}
    \caption{Comparison of execution accuracy and average token length for the text-to-SQL task across different retrieval methods. ATR achieves higher accuracy with fewer tokens compared to the best-performing top-$k$ approach.}
    \label{fig:Downstream Tasks Efficiency Comparison}
\end{figure*}

\paragraph{Baseline}
% We compare ATR with the two bi-encoder baselines—Contriever and UAE-Large-V1—and with the reranking method JAR~\citep{chen-etal-2024-table}, which explicitly encodes table joinability. 
% Contriever and UAE-Large-V1 embed the query and each table independently, flattening the table into text and ranking candidates by cosine similarity between their vector representations. 
% Both ATR and JAR adopt a two-stage reranking pipeline: a bi-encoder first retrieves the top 50 tables, and then the reranker reranks the tables, followed by standard approaches~\citep{glass-etal-2022-re2g, sun-etal-2023-chatgpt,qin-etal-2024-large}.
We compare ATR with two bi-encoder baselines, Contriever and UAE-Large-V1, and with the reranking method JAR~\citep{chen-etal-2024-table}, which encodes table joinability.
Contriever and UAE-Large-V1 embed the query and each table independently by flattening tables into text, and rank tables based on cosine similarity between their embedding vectors.
Both ATR and JAR utilize a two-stage retrieval pipeline, where a bi-encoder first retrieves the top 50 candidate tables, and a reranker subsequently refines their order, following standard practice~\citep{glass-etal-2022-re2g, sun-etal-2023-chatgpt, qin-etal-2024-large,wasserman-etal-2025-docrerank}.
Additionally, we include LLM-based retrieval methods: RankZephyr~\citep{pradeep2023rankzephyreffectiverobustzeroshot}, an open-source LLM specialized in listwise zero-shot reranking, and Murre~\citep{zhang-etal-2025-murre}, a multi-hop retrieval method based on LLM that iteratively retrieves tables.

\subsection{Table Retrieval Performance}
Across all tested values of $k$, ATR outperforms most of the top-$k$ baselines on the in-domain datasets (Spider and BIRD) and out-of-domain dataset (Spider 2.0), as shown in Table~\ref{Experiments:Main Result}. 
% On Spider, ATR improves F1 by more than 15.0\% over all the comparison targets, showing much higher precision than others. 
On Spider, ATR achieves substantial gains over all the comparison targets, improving precision by more than 20.0\% and F1 by over 15.0\%.
Compared to other baselines, it performs almost the best on precision and recall, and achieves the highest F1 score on BIRD.
Although no query in Spider or BIRD requires more than five tables and the baselines already use conservative $k$ values, our method still surpasses them.
The performance improvement of recall becomes more evident on Spider 2.0, where the number of ground-truth tables varies from one to hundreds. 
% With $k=10$, ATR improves complete recall by 19.6\% over Contriever and 23.0\% over JAR.  
ATR shows a characteristic that the recall and complete recall scores are at least 10.0\% higher than other models.
These findings confirm that the top-$k$ retrieval strategy cannot be generalized across queries requiring varying numbers of tables, whereas ATR demonstrates robustness to such variations. 

\subsection{End-to-end Performance}
We evaluate end-to-end effectiveness through the text-to-SQL task.
To set an upper bound for retrieval-based approaches, we report an \emph{Oracle} setting where the generator receives only the ground-truth tables.
As described in Table~\ref{tab:end-to-end_performance}, ATR shows the best execution accuracy across all three datasets and evaluated generators. 
With Qwen2.5-Coder-32B and $k=10$, ATR improves on JAR by 3.7\% on Spider and 1.7\% on BIRD when using UAE-Large-V1 as the bi-encoder; the performance gains increase to 5.2\% and 2.5\% when the 7B model is used, respectively.
Compared to Oracle, ATR has a difference only 2.6\% and 3.6\% on Spider and BIRD with Llama-3.1-70B, respectively. 
On Spider 2.0, there is a difference of only 1.4\% with Qwen2.5-Coder-7B.
% For Gemma-3-27B-IT, the performance gap with the Oracle setting is minimal, with only 0.6\% on Spider and 1.2\% on BIRD.
% Regarding Gemma-3-4B-IT, the gap is only 0.3\% on Spider 2.0.
These results confirm earlier findings that stronger retrieval performance translates into higher downstream accuracy~\citep{kothyari-etal-2023-crush4sql,chen-etal-2024-table}. 

% With Qwen2.5-Coder-32B and $k=10$, ATR improves on JAR by 2.3\% on Spider and 0.3\% on BIRD; the performance gains increase to 3.1\% and 2.0\%, respectively, when the small 7B model is used.
% On the more demanding Spider 2.0, ATR surpasses Contriever and JAR by 1.6\% and 1.3\%. 
% The trend is consistent for Llama-3.1-70B: ATR outperforms JAR by 1.0\% on Spider and 2.4\% on BIRD with $k=10$. 

\section{Analysis}
% This section provides an in-depth analysis of ATR, focusing on token efficiency, robustness to varying numbers of required ground-truth tables, and comparisons with state-of-the-art adaptive document retrieval methods.
% Additionally, we provide ablation and statistical analyses to empirically validate the effectiveness of table representations.

\begin{figure}[t]
    \centering
    \includegraphics[width=\columnwidth,height=3.3cm]{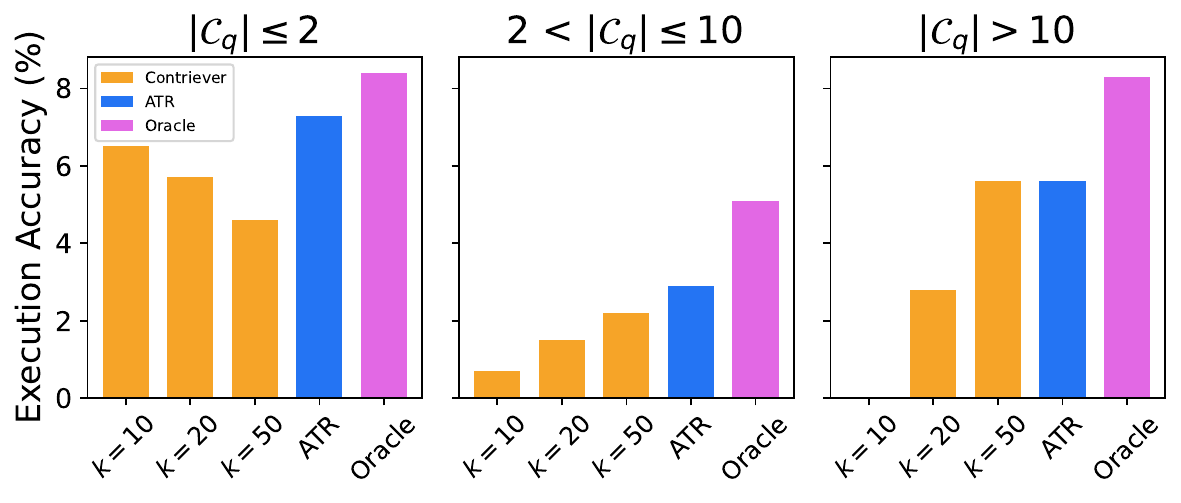}
    \caption{Execution accuracy on the Spider 2.0 dataset across varying numbers of required tables. $|\mathcal{C}_q|$ denotes the number of tables required per query.}
    \label{Analysis:performance_variation}
\end{figure}

\begin{figure*}[t]
    \centering
    % ---------------- [1] figure ----------------
    \includegraphics[width=\textwidth, height=7cm]{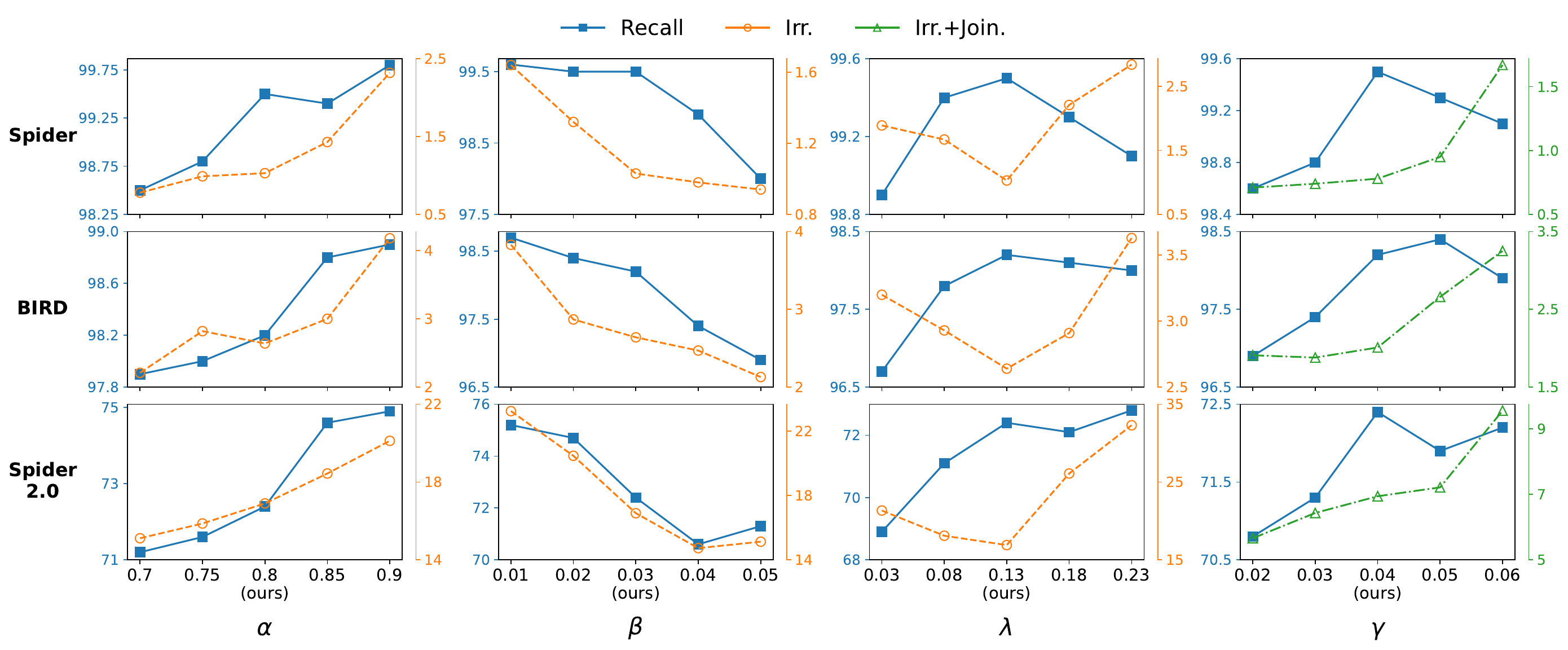}
    \caption{Analysis of training loss hyper-parameters. Irr. indicates the number of retrieved tables irrelevant to the query, and Irr.+Join. represents the retrieved tables that are irrelevant but can be joined with the ground-truth tables.}
    \label{Analysis:sensitivity_hyperparameters}

    \vspace{0.8cm}

    % ---------------- [2] table ----------------
    \centering
    \begin{tabular}{cccccccccc}
    \hline
    & \multicolumn{3}{c}{\textbf{Spider}} & \multicolumn{3}{c}{\textbf{BIRD}} & \multicolumn{3}{c}{\textbf{Spider 2.0}} \\ \hline
    \textbf{$\bm{W, R}$} & 50, N/A & 20, 15 & 10, 6 & 50, N/A & 20, 15 & 10, 6 & 50, N/A & 10, 5 & 6, 4 \\ \hline
    \textbf{Time} & 0.12 & 0.14 & 0.17 & 0.18 & 0.21 & 0.26 & 0.60 & 0.41 & 0.45 \\
    \textbf{Length} & 1164.4 & 471.7 & 237.5 & 1592.7 & 617.9 & 324.0 & 3982.8 & 846.2 & 532.8 \\
    \textbf{\# Inference} & 1 & 7 & 11 & 1 & 7 & 11 & 1 & 9 & 23 \\ \hline
    \end{tabular}
    
    \captionof{table}{Analysis of sliding window hyper-parameters $(W, R)$, where $W$ denotes the window size and $R$ is the retention size. The metrics include average inference time in seconds (Time), average input token length (Length), and number of inferences (\# Inference).}
    \label{tab:sliding analysis}
    
\end{figure*}

\subsection{Efficiency and Accuracy Improvement}
Retrieving too few tables omits evidence and degrades performance, whereas retrieving too many injects noise and inflates inference cost.
We analyze how many tables ATR and top-$k$ methods retrieve compared to the Oracle, and how this affects downstream performance.
For the top-$k$ methods, we vary $k$ from 1 to 10 for Spider and BIRD, and from 5 to 50 with intervals of 5 for Spider 2.0.

Figure~\ref{fig:Downstream Tasks Efficiency Comparison} shows execution accuracy against the average number of input tokens used to generate SQL queries. 
ATR achieves higher execution accuracy with fewer tokens than top-$k$ baselines on all three datasets. 
Specifically, ATR uses 430 fewer tokens than the best-performing top-$k$ ($k=8$) on Spider, and 522 fewer tokens than the best top-$k$ ($k=10$) on BIRD. 
When compared to top-$k$ baselines with similar token budget, ATR demonstrates higher execution accuracy—improving by 5.2\% on Spider ($k=3$) and 3.5\% on BIRD ($k=5$).
The gap widens on the out-of-domain Spider 2.0 dataset, where ATR narrows the gap toward the Oracle, which is the upper bound.
These results demonstrate that ATR is a robust method that retrieves ground-truth tables accurately while minimizing the retrieval of irrelevant tables. 

\subsection{Performance by Number of Required Tables}
The fixed $k$ retrieval strategy inherently suffers from trade-offs; either failing to retrieve necessary tables or retrieving irrelevant ones.
To illustrate this trade-off and demonstrate the robustness of our method, we analyze the downstream task performance by categorizing the Spider 2.0 queries into three groups based on the number of ground-truth tables: two or fewer, between three and ten, and more than ten. 

Figure~\ref{Analysis:performance_variation} illustrates the limitations of a fixed $k$ approach. 
Top-$k$ retrieval performs best on queries that require two or fewer tables when using a smaller retrieval count ($k=10$), but its performance collapses for queries that need more than ten tables.
Conversely, using a larger retrieval count ($k=50$) enhances performance for queries that require more than ten tables, but it falls for queries that require two or fewer tables because of noise from irrelevant tables. 
ATR addresses this trade-off, consistently outperforming baselines across queries with various table requirements.

\begin{table}[t]
\centering
\setlength{\tabcolsep}{1mm} 
\begin{adjustbox}{width=1\columnwidth}
\begin{tabular}{lcccccc}
\toprule
\textbf{} & \multicolumn{3}{c}{\textbf{Spider}} & \multicolumn{3}{c}{\textbf{BIRD}} \\ \cline{2-7}

 & \textbf{R} & \textbf{Acc.} & \textbf{Time($\downarrow$)} & \textbf{R} & \textbf{Acc.} & \textbf{Time($\downarrow$)} \\ \hline
 
\textbf{{Iter-RetGen}} & \underline{98.8} & {\textbf{71.7}} & 8.9 & \underline{96.5} & \underline{52.7} & 14.4 \\

\textbf{{FLARE}} & 89.0 & 63.2 & \underline{4.0} & 78.3 & 43.1 & \underline{5.7} \\ 

\textbf{{Adaptive-RAG}} & 86.0 & 62.8 & 6.3 & 89.8 & 50.7 & 13.2 \\ 
\hline
\textbf{ATR (Ours)} & \textbf{99.5} & \underline{71.5} & \textbf{2.2} & \textbf{98.2} & \textbf{53.3} & \textbf{3.8} \\
\bottomrule
\end{tabular}
\end{adjustbox}
\caption{ATR outperforms existing adaptive document retrieval strategies in both efficiency and accuracy. Acc. and Time denote execution accuracy and the average end-to-end inference time in seconds, respectively.}
\label{tab:adaptive_retrieval}
\end{table}

\subsection{Comparison with Adaptive Document Retrieval}
Recent adaptive document retrieval methods adjust the retrieval strategy based on query complexity. 
FLARE~\citep{jiang-etal-2023-active} triggers additional retrieval whenever the generator outputs low-confidence tokens, and Adaptive-RAG~\citep{jeong2024adaptive} trains a classifier to determine the number of retrieval iterations based on query complexity.
Iter-RetGen~\citep{shao2023enhancing} serves as an iterative baseline, retrieving documents repeatedly for a fixed number of iterations.
To compare ATR with these adaptive document retrieval methods, we conduct experiments on the Spider and BIRD datasets with Qwen2.5-Coder-32B. 
We evaluate recall, text-to-SQL execution accuracy, and end-to-end latency in seconds.
To ensure fairness, we train the Adaptive-RAG classifier on the Spider and BIRD training splits. 

Table~\ref{tab:adaptive_retrieval} shows that ATR consistently outperforms adaptive document retrieval baselines in both retrieval and end-to-end tasks. 
Compared to Adaptive-RAG, ATR improves execution accuracy by 8.7\% on Spider and 2.6\% on BIRD, simultaneously reducing processing times by 4.1 and 9.4 seconds, respectively.
Against Iter-RetGen, ATR reduces the inference time by 6.7 seconds on Spider and 10.6 seconds on BIRD, while achieving comparable execution accuracy. 
These results confirm the effectiveness of ATR through improved table representation learning and the efficiency of operating without iterative LLM retrieval interactions.

\subsection{Hyper-parameter Analysis}
\label{subsec:hyper_parameter_analysis}
We analyze how retrieval behavior changes with respect to hyper-parameters controlling each loss component.
$\alpha$ and $\beta$ control the margin between relevant and irrelevant tables based on the threshold token.
As shown in Figure~\ref{Analysis:sensitivity_hyperparameters}, increasing $\alpha$ and decreasing $\beta$ tend to increase the number of tables included in retrieval results, consequently improving recall but also increasing the number of irrelevant tables.
Similarly, $\lambda$ and $\gamma$ significantly impact both recall and the number of retrieved tables. 
Notably, as $\gamma$ increases, retrieval results include more tables that are irrelevant but joinable with the ground-truth tables.
These observations align closely with our design of the loss functions.
Furthermore, an ANOVA test confirms that the threshold token forms a statistically significant decision boundary ($p < 0.05$) separating relevant and irrelevant tables, as detailed in Appendix~\ref{app:anova}.

ATR performs re-ranking on the top 50 tables initially retrieved by bi-encoder, varying the window and retention sizes. 
Depending on the window size and retention size, inference time can vary due to changes in the average input token length and the number of required inferences.
We present in Table~\ref{tab:sliding analysis} how the input length, number of inferences, and inference time vary according to the window and retention sizes.
For Spider and BIRD datasets, tables generally have shorter schemas, resulting in shorter token lengths; thus, inference time is primarily influenced by the number of inferences.
Conversely, for Spider 2.0, which contains larger individual tables with longer schemas, inference time is predominantly determined by the window size, directly affecting input token length per inference.

\begin{table}[t!]
\centering
\setlength{\tabcolsep}{1.2mm} 
\resizebox{\columnwidth}{!}{%
\begin{tabular}{lcccccc}
\hline
                       & \multicolumn{2}{c}{\textbf{Spider}}             & \multicolumn{2}{c}{\textbf{BIRD}}               & \multicolumn{2}{c}{\textbf{Spider 2.0}}    \\ \cline{2-7}
                       & \textbf{R}      & \textbf{CR} & \textbf{R}      & \textbf{CR} & \textbf{R} & \textbf{CR} \\ \hline
\textbf{ATR}          & { \textbf{99.5}} & { \textbf{99.2}}     & {\textbf{98.2}} & { \textbf{96.0}}     & {\textbf{72.4}}     & { \textbf{64.4}}              \\ \hline
$-$ (1) BCE         & \underline{99.0}                & \underline{98.7}              & \underline{97.5}          & \underline{95.8}              & \underline{68.2}     & \underline{60.8}              \\
$-$ (2) Contrastive & \underline{99.0}          & 98.4                    & 97.4                & 95.2                    & 69.1           & 60.1                    \\ 
$-$ (1) \& (2)        & 96.4                & 94.4                    & 91.8                & 85.7                    & 67.7                    & 58.2                    \\ 
\hline
\end{tabular}%
}
\caption{Ablation study on table representation losses.}
% \caption{Ablation study on training objectives for table representation. Both loss functions contribute significantly to the retrieval performance of ATR.}
\label{tab:ablation}
% \end{adjustbox}
\end{table}

\subsection{Ablation Study}
ATR is trained with two auxiliary objectives: a BCE loss for relevance calibration and a contrastive loss for semantic grouping.
To evaluate each loss component, we train separate models by removing each objective individually.

As shown in Table~\ref{tab:ablation}, removing the BCE loss results in both lower recall and complete recall, confirming that explicit query-table alignment is crucial for distinguishing relevant tables from irrelevant ones. 
Similarly, removing contrastive loss reduces retrieval performance, indicating that inter-table joinability enhances the discriminative quality of table embeddings, thereby improving retrieval accuracy. 
These findings indicate that both losses are essential for learning robust and discriminative table representations, which in turn enhance the retrieval performance of ATR over existing table retrieval methods.

\section{Conclusion}
In this work, we address the inherent limitations of conventional table retrieval methods that rely on a fixed $k$ retrieval strategy.
The rigidity in the fixed $k$ often degrades downstream task performance and efficiency by retrieving unnecessary tables or failing to retrieve tables required for accurate reasoning. 
To mitigate these problems, we propose ATR, an adaptive table retrieval method that dynamically adjusts the number of retrieved tables based on the query. 
ATR leverages adaptive thresholding to determine the optimal number of tables required for each query. 
Furthermore, ATR adopts relevance calibration and semantic grouping loss to effectively learn table representations by capturing query-to-table and inter-table relationships. 
Extensive experiments demonstrate that ATR consistently outperforms top-$k$ retrieval methods, demonstrating superior retrieval performance, downstream accuracy, and inference efficiency. 
These results confirm that ATR can be a practical solution suitable for large-scale database retrieval applications.

\section*{Limitations}
Although ATR demonstrates substantial improvements in both retrieval and downstream execution accuracy, there is room for further improvement.
ATR currently targets structured tabular data exclusively, and extending its adaptive retrieval strategy to handle other data modalities or mixed data types remains an open research challenge.
Nevertheless, ATR provides a robust and efficient adaptive retrieval framework without direct interaction with generative LLMs. 
By explicitly learning effective table representations and efficiently managing retrieval through adaptive thresholding, ATR establishes a strong foundation for future studies aiming to generalize retrieval-augmented generation across various data types and broader application contexts.

\bibliography{custom}

\appendix

\clearpage

\section{Implementation Details}
\label{app:environment}
We train ATR using two NVIDIA RTX A6000 GPUs, each equipped with 48GB of memory.
Training is conducted for three epochs with a batch size of 64 and a learning rate of \(3e^{-5}\).
The maximum token length of ATR is set to 8,192, matching the input length constraints of ModernBERT-large.
For the adaptive thresholding loss, the hyper-parameters are set as \(\alpha = 0.8\) and \(\beta = 0.03\).
For relevance calibration and semantic grouping loss, we select \(\lambda = 0.13\) and \(\gamma = 0.04\), respectively.
The window size is set to 20 for the Spider and BIRD datasets, and 10 for Spider 2.0-Lite.
The retention size is set to 15 for Spider and BIRD, and 5 for Spider 2.0-Lite.

For our experiments, we utilize the following models:
We employ ModernBERT-large~\citep{modernbert}, specifically the \texttt{answerdotai/ModernBERT-large}, as the backbone of ATR.
For table embedding, we employ Contriever~\citep{izacard2021unsupervised}, using the \texttt{facebook/contriever\-msmarco}, and UAE-Large-V1~\citep{li-li-2024-aoe} with the \texttt{WhereIsAI/UAE-Large-V1}. 
For LLM-based retrieval, we leverage RankZephyr~\citep{pradeep2023rankzephyreffectiverobustzeroshot}, specifically the \texttt{castorini/rank\_zephyr\_7b\_v1\_full}, and SGPT~\citep{muennighoff2022sgpt} from the \texttt{Muennighoff/SGPT-5.8B-weightedmean\-msmarco-specb-bitfit}.

\section{Dataset Pre-processing}
\label{app:dataset_processing}

ATR is trained on the union of the Spider and BIRD training splits. 
For each query, we first use Contriever to retrieve the top 100 tables and then split this list in half: the higher-ranked segments that rank from 1 to 50 and the lower-ranked segments that rank from 51 to 100.
We pair each segment with the query to create two training examples: one that is likely to contain relevant tables and one that is likely not. 
This contrastive environment makes ATR learn how to operate when the candidate set contains no relevant tables, reducing false positives at inference time. 
We split this dataset into training and validation sets at a ratio of 85\% and 15\% and the best checkpoint is selected by validation performance. 

ATR adopts a semantic grouping loss to effectively learn table representations by capturing table-to-table relationships.
To achieve this, we leverage joinability information between tables.
Specifically, we identify joinable table groups by performing syntactic analysis on the publicly available database schemas from the Spider and BIRD training datasets.
We exclude training samples for which joinability cannot be determined from the given database corpus.
Additionally, we remove tables that exceed the maximum input token length of 512 tokens, consistent with the constraints of the bi-encoders used in our experiments, along with queries requiring these tables as ground truths.
Furthermore, we exclude cases from Spider 2.0-Lite where tables labeled as ground truths are not present in the corresponding databases.
Dataset statistics are summarized in Table~\ref{tab:datasets}.

\begin{table}[t]
\centering
\begin{adjustbox}{width=1\columnwidth}
\begin{tabular}{rcccc}
\hline
\multicolumn{1}{l}{\textbf{Dataset}}    & \textbf{\#Q} & \textbf{\#DB} & \textbf{\#T} & \textbf{Min/Max} \\ \hline
\multicolumn{1}{l}{\textbf{Spider}}     &                    &               &  &                \\
Train                          & 6,989              & 140           & 737  &1 / 5              \\
Eval                           & 1,034              & 20            & 81   &1 / 4            \\ \hline
\multicolumn{1}{l}{\textbf{BIRD}}       &                    &               &       &           \\
Train                          & 9,198              & 69            & 522  &1 / 4            \\
Eval                           & 1,534              & 11            & 75   &1 / 4            \\ \hline
\multicolumn{1}{l}{\textbf{Spider 2.0}} &                    &               &  &                \\
Eval                           & 435                & 155           & 6,321 &1 / 366           \\ \hline     
\end{tabular}
\end{adjustbox}
\caption{Data statistics. Number of queries (\#Q), databases (\#DB), tables (\#T), and required tables (Min/Max). Evaluations on Spider and BIRD use development sets.}
\label{tab:datasets}
\end{table}

\begin{table}[t]
\centering
\setlength{\tabcolsep}{1.3mm} 
\begin{adjustbox}{width=1\columnwidth}
\begin{tabular}{lccccccc}
\toprule
                &      & \multicolumn{2}{c}{\textbf{Spider}}                                                     & \multicolumn{2}{c}{\textbf{BIRD}}                                                         & \multicolumn{2}{c}{\textbf{Spider 2.0}}                                                   \\ 
                \cmidrule{3-8}
                &   $k$   & \textbf{R}    & \textbf{\begin{tabular}[c]{@{}c@{}}CR\end{tabular}} & \textbf{R}      & \textbf{\begin{tabular}[c]{@{}c@{}}CR\end{tabular}} & \textbf{R}      & \textbf{\begin{tabular}[c]{@{}c@{}}CR\end{tabular}} \\ \toprule
\textbf{OpenAI} & 3  & 96.8              & 93.5                                                              & 85.8                & 72.1                                                              & 40.7                & 28.2                                                              \\
\textbf{}       & 5  & \underline{99.7}              & 99.4                                                              & 92.8                & 85.5                                                              & 50.0                & 36.7                                                              \\
\textbf{}       & 10 & { \textbf{100}} & { \textbf{100}}                                                 & \underline{ 98.4} & \underline{ 96.7}                                               & \underline{ 62.8}          & \underline{ 49.8}                                                        \\ \midrule
\textbf{ATR (Ours)}   &      & 99.6        & \underline{ 99.6}                                                        & {\textbf{99.5}}          & {\textbf{99.2}}                                                        & {\textbf{79.5}} & {\textbf{70.2}}                                               \\ 
\bottomrule
\end{tabular}
\end{adjustbox}
\caption{Evaluation of retrieval performance comparing ATR and a proprietary embedding model. OpenAI indicates text-embedding-3-large.}
\label{tab:proprietary embed}
\end{table}

\begin{table*}[t]
\centering
\resizebox{\textwidth}{!}{
\begin{tabular}{lc cccc cccc cccc}
\toprule
\multirow{2}{*}{} & & \multicolumn{4}{c}{\textbf{Spider}} & \multicolumn{4}{c}{\textbf{BIRD}} & \multicolumn{4}{c}{\textbf{Spider 2.0}} \\
\cmidrule(lr){3-6} \cmidrule(lr){7-10} \cmidrule(lr){11-14}
 & \textbf{k} & P & R & CR & F1 & P & R & CR & F1 & P & R & CR & F1 \\
\midrule
RankGPT-4o-mini & 3 & 48.2 & 96.8 & 93.5 & 62.1 & \textit{54.5} & 88.0 & 76.2 & 65.6 & 33.2 & 48.5 & 33.8 & 35.2 \\
RankGPT-4o-mini & 5 & 29.9 & 98.9 & 97.9 & 44.5 & 35.0 & 92.7 & 84.6 & 49.7 & 24.6 & 56.1 & 42.1 & 30.5 \\
RankGPT-4o-mini & 10 & 15.1 & 99.3 & 99.0 & 25.5 & 18.7 & 97.5 & 94.6 & 30.9 & 15.3 & 64.0 & 52.2 & 21.9 \\
\midrule
RankGPT-5-mini & 3 & 48.9 & 98.2 & 95.5 & 63.1 & \textbf{57.3} & 91.6 & 81.6 & \textbf{66.3} & \textbf{36.6} & 53.6 & 37.5 & \textbf{39.4} \\
RankGPT-5-mini & 5 & 30.0 & 99.5 & 98.9 & 44.8 & 36.4 & 95.7 & 90.0 & 51.5 & 25.8 & 59.3 & 44.1 & 32.4 \\
RankGPT-5-mini & 10 & 15.1 & \textbf{100.0} & \textbf{100.0} & 25.8 & 18.9 & 98.0 & 96.0 & 31.4 & 15.8 & 66.1 & 54.2 & 22.7 \\
\midrule
RankQwen3-32B & 3 & 48.4 & 97.4 & 94.2 & 62.5 & 53.1 & 88.7 & 76.4 & 65.5 & \textit{33.3} & 48.4 & 32.6 & \textit{35.4} \\
RankQwen3-32B & 5 & 30.1 & 99.5 & 98.9 & 44.8 & 35.8 & 94.6 & 87.6 & 50.8 & 24.3 & 55.5 & 40.0 & 30.2 \\
RankQwen3-32B & 10 & 15.1 & 99.4 & 99.0 & 25.6 & 18.9 & 98.0 & \textit{96.4} & 31.2 & 15.3 & 64.0 & 52.0 & 22.0 \\
\midrule
ATR (w. Contriever) &  & \textbf{69.6} & 99.5 & 99.2 & \textbf{78.3} & 54.0 & \textit{98.2} & 96.0 & \textit{65.8} & 21.9 & \textit{72.4} & \textit{64.4} & 27.8 \\
ATR (w. UAE) &  & \textit{69.3} & \textit{99.6} & \textit{99.4} & \textit{78.1} & 52.8 & \textbf{98.6} & \textbf{97.1} & 65.1 & 19.9 & \textbf{75.4} & \textbf{68.7} & 26.7 \\
\bottomrule
\end{tabular}
}
\caption{Performance comparison with LLM-based retrievers and rerankers. \textbf{Bold} indicates the highest score, and \textit{italic} indicates the second-highest score.}
\label{tab:llm_reranker}
\end{table*}

\begin{table*}[t]
\centering
\resizebox{\textwidth}{!}{%
\begin{tabular}{lc cccc cccc cccc}
\toprule
 &  & \multicolumn{4}{c}{\textbf{Spider}} & \multicolumn{4}{c}{\textbf{BIRD}} & \multicolumn{4}{c}{\textbf{Spider 2.0}} \\ \cmidrule(lr){3-6} \cmidrule(lr){7-10} \cmidrule(lr){11-14}
 & \textit{k} & \textbf{Qwen3} & \textbf{DeepSeek-V2} & \textbf{GPT-5} & \textbf{Gemini-2.5} & \textbf{Qwen3} & \textbf{DeepSeek-V2} & \textbf{GPT-5} & \textbf{Gemini-2.5} & \textbf{Qwen3} & \textbf{DeepSeek-V2} & \textbf{GPT-5} & \textbf{Gemini-2.5} \\ \hline
\multicolumn{14}{l}{\textit{\textbf{Encoder-based}}} \\ \hline
Contriever & 3 & 71.7 & 61.7 & 60.4 & 73.7 & 45.2 & 29.9 & 35.1 & 48.8 & 3.2 & 1.8 & 3.2 & 2.5 \\
 & 5 & 74.2 & 61.9 & 63.4 & 75.0 & 50.3 & 32.3 & 37.5 & 54.0 & 3.2 & 1.6 & 5.1 & 3.9 \\
 & 10 & 75.9 & 62.3 & 64.8 & 76.5 & 53.7 & 36.8 & 40.4 & 58.1 & 3.7 & 1.6 & 5.8 & 4.4 \\
UAE & 3 & 71.9 & 61.5 & 60.6 & 73.3 & 46.1 & 31.8 & 34.2 & 50.5 & 3.7 & 1.6 & 4.6 & 2.5 \\
 & 5 & 75.9 & 62.4 & 64.0 & 76.3 & 49.7 & 33.6 & 37.6 & 54.8 & 3.9 & \underline{2.8} & 5.3 & 4.6 \\
 & 10 & 76.2 & 63.8 & 65.1 & 76.9 & 54.0 & 38.3 & 41.3 & 59.3 & 4.4 & 1.8 & 7.6 & 4.6 \\ \hline
\multicolumn{14}{l}{\textit{\textbf{LLM-based}}} \\ \hline
RankZephyr & 3 & 71.3 & 62.4 & 60.1 & 73.2 & 45.1 & 29.7 & 36.0 & 48.6 & 3.2 & 2.1 & 5.1 & 4.1 \\
 & 5 & 74.3 & 61.3 & 64.9 & 75.4 & 51.2 & 33.1 & 37.9 & 54.1 & 3.5 & 0.9 & 4.6 & 3.5 \\
 & 10 & 76.2 & 63.2 & 65.0 & 76.9 & \underline{54.6} & 37.0 & 40.0 & 58.7 & 3.7 & 1.8 & 7.4 & 5.1 \\
Murre & 3 & 71.3 & 62.4 & 60.7 & 73.9 & 49.3 & 30.1 & 36.1 & 50.3 & 2.5 & 2.3 & 4.1 & 3.5 \\
 & 5 & 74.1 & 63.6 & 62.8 & 74.6 & 50.5 & 32.5 & 37.0 & 52.5 & 3.9 & 2.3 & 5.8 & 5.3 \\
 & 10 & 75.9 & 64.4 & 64.3 & 76.2 & 53.6 & 37.3 & 41.3 & 58.7 & 4.1 & 2.1 & 7.4 & 5.5 \\ \hline
\multicolumn{14}{l}{\textit{\textbf{Reranker-based}}} \\ \hline
JAR  & 3 & 74.5 & 63.8 & 63.2 & 74.8 & 49.3 & 36.3 & 37.1 & 55.3 & 2.3 & 2.1 & 4.8 & 3.2 \\
(w. Contriever) & 5 & 74.6 & 61.9 & 64.5 & 76.4 & 52.4 & 37.0 & 37.7 & 56.5 & 3.0 & 1.6 & 5.3 & 4.6 \\
 & 10 & 75.6 & 63.1 & 66.2 & 76.5 & 54.0 & 38.2 & 39.4 & 58.1 & 3.9 & 2.3 & 5.5 & 4.4 \\
JAR  & 3 & 74.6 & 62.9 & 65.4 & 75.8 & 49.6 & 36.3 & 37.2 & 52.9 & 3.0 & 2.5 & 4.4 & 4.1 \\
(w. UAE) & 5 & 76.2 & 63.1 & 67.1 & 77.4 & 50.6 & 36.8 & 38.0 & 54.3 & 3.2 & 2.3 & 5.8 & 4.6 \\
 & 10 & 75.6 & 64.7 & \underline{67.3} & \underline{78.1} & 53.6 & 38.0 & 41.1 & 57.7 & 3.0 & 2.5 & 6.9 & 5.5 \\ \hline
\multicolumn{14}{l}{\textit{\textbf{Ours}}} \\ \hline
ATR &  & \underline{76.3} & \textbf{66.4} & 65.5 & 77.9 & \textbf{55.0} & \underline{38.5} & \underline{41.6} & \underline{59.4} & \underline{4.6} & \textbf{3.0} & \textbf{9.0} & \underline{5.8} \\
(w. Contriever) & & & & & & & & & & & & & \\
ATR &  & \textbf{76.7} & \underline{65.8} & \textbf{67.8} & \textbf{79.4} & 54.3 & \textbf{39.6} & \textbf{41.9} & \textbf{59.5} & \textbf{4.8} & 2.5 & \underline{8.5} & \textbf{6.7} \\ 
(w. UAE) & & & & & & & & & & & & & \\
\hline
\textit{\textbf{Oracle}} &  & 79.0 & 71.4 & 68.3 & 79.7 & 59.5 & 44.9 & 43.1 & 62.8 & 9.0 & 5.1 & 11.5 & 8.9 \\ \bottomrule
\end{tabular}%
}
\caption{Text-to-SQL execution accuracy comparison for the various models, including open source and proprietary LLMs, across different table retrieval methods.}
\label{tab:additional_end-to-end_performance}
\end{table*}

% \begin{figure*}[t]
%     \centering
%     \includegraphics[width=\textwidth, height=7cm]{latex/images/hyper_parameter_analysis_color.pdf}
%     \caption{Analysis of training loss hyper-parameters. Irr. indicates the number of retrieved tables irrelevant to the query, and Irr.+Join. represents the retrieved tables that are irrelevant but can be joined with the ground-truth tables.}
%     \label{Analysis:sensitivity_hyperparameters}

%     \vspace{0.5cm} 

%     \centering
%     \begin{tabular}{cccccccccc}
%     \hline
%     & \multicolumn{3}{c}{\textbf{Spider}} & \multicolumn{3}{c}{\textbf{BIRD}} & \multicolumn{3}{c}{\textbf{Spider 2.0}} \\ \hline
%     \textbf{$\bm{W, R}$} & 50, N/A & 20, 15 & 10, 6 & 50, N/A & 20, 15 & 10, 6 & 50, N/A & 10, 5 & 6, 4 \\ \hline
%     \textbf{Time} & 0.12 & 0.14 & 0.17 & 0.18 & 0.21 & 0.26 & 0.60 & 0.41 & 0.45 \\
%     \textbf{Length} & 1164.4 & 471.7 & 237.5 & 1592.7 & 617.9 & 324.0 & 3982.8 & 846.2 & 532.8 \\
%     \textbf{\# Inference} & 1 & 7 & 11 & 1 & 7 & 11 & 1 & 9 & 23 \\ \hline
%     \end{tabular}
    
%     \captionof{table}{Analysis of sliding window hyper-parameters $(W, R)$, where $W$ denotes the window size and $R$ is the retention size. The metrics include average inference time in seconds (Time), average input token length (Length), and number of inferences (\# Inference).}
%     \label{tab:sliding analysis}
    
% \end{figure*}

\section{Effectiveness of ATR Leveraging an Advanced Embedding Model}
\label{app:additional_retrieval_results}
We evaluate the effectiveness of ATR when leveraging an advanced embedding model to potentially enhance retrieval performance.
Our experiments use text-embedding-3-large, a state-of-the-art proprietary embedding model, and results are shown in Table~\ref{tab:proprietary embed}.
ATR achieves high complete recall, obtaining 99.2\% on the BIRD and 70.2\% on Spider 2.0.
These results highlight ATR’s robustness and effectiveness, demonstrating that it can achieve strong retrieval performance when combined with advanced embedding models.

\section{Comparison with Combined LLM-based Pipelines}
\label{sec:llm_pipelines}

To provide a comprehensive evaluation against state-of-the-art LLMs, we expand our baselines to include combined LLM-based retriever-and-reranker pipelines.
Specifically, we evaluate a pipeline that utilizes powerful LLMs, including GPT-4o-mini, GPT-5-mini, and Qwen3-32B, to rerank the initially retrieved tables at various fixed top-$k$ cutoffs.

As shown in Table~\ref{tab:llm_reranker}, ATR achieves comparable or superior performance across all datasets.
While LLM-based rerankers inherently suffer from a precision-recall trade-off dictated by a fixed top-$k$ value, ATR balances this without requiring a pre-defined cutoff.
Crucially, LLMs impose substantial computational overhead and API costs.
ATR, on the other hand, uses a lightweight bi-encoder to deliver highly efficient and practical retrieval suitable for real-world enterprise workflows.

\section{Additional End-to-end Performance Results}
\label{app:additional_end-to-end_results}
To further verify the robustness of the end-to-end performance of ATR, we conduct additional experiments on SQL generation utilizing Qwen3-Coder-30B-A3B-Instruct~\citep{qwen3technicalreport}, DeepSeek-Coder-V2-Lite-Instruct~\citep{zhu2024deepseek}, GPT-5-mini~\citep{openai2025gpt51}, and Gemini-2.5-flash~\citep{comanici2025gemini}. 
As shown in Table~\ref{tab:additional_end-to-end_performance}, ATR consistently outperforms all baselines regardless of whether the generator is an open-source or proprietary model.
Notably, using Gemini-2.5-flash, ATR narrows the performance gap with the Oracle setting to just 0.3\% on Spider and 3.3\% on BIRD. 
Furthermore, with Qwen3-Coder-30B-A3B-Instruct and DeepSeek-Coder-V2-Lite-Instruct, ATR maintains gaps of 4.2\% and 2.1\% in the Oracle setting on Spider 2.0, while substantially outperforming all baselines.
% Furthermore, on Spider 2.0, GPT-5-mini with ATR achieves an execution accuracy of 9.0\%, surpassing even the Oracle performance of Gemini-2.5-flash. 
These results demonstrate that ATR effectively retrieves the optimal set of tables, enabling advanced LLMs to fully leverage their reasoning capabilities.

\section{Statistical Analysis of Table Representations}
\label{app:anova}
ATR assigns logits on tokens $T_{th}$ for the threshold and $T_{i}$ for the table representation.
We use analysis of variance (ANOVA) to investigate differences between group means within relevant tables, irrelevant tables, and a threshold. 
Most of the variance is explained by the difference of group means on Spider, revealing large effects ($\eta^{2} \approx 0.95$) with significant $p$-values ($p < 0.05$). 
On the BIRD and Spider 2.0 datasets, ANOVA reveals significant effects ($\eta^{2} \approx 0.86, 0.15$) with significant $p$-values ($p < 0.05$).
A pairwise Tukey post-hoc test reveals a significant difference ($p < 0.05$ for all the pairs) between relevant tables, irrelevant tables, and the threshold for the three datasets.
These results confirm that ATR robustly differentiates between relevant and irrelevant tables, with the adaptive threshold serving as a clear decision boundary that guides accurate table selection for each query.

\begin{table}[ht]
\centering
\resizebox{\columnwidth}{!}{
\begin{tabular}{ll ccc}
\toprule
\textbf{Dataset} & \textbf{Metric} & \textbf{Sliding Window (Ours)} & \textbf{No Sliding Window} & \textbf{Reduction} \\
\midrule
\multirow{2}{*}{Spider} & Avg Peak & 30.62 MB & 63.98 MB & 52.1\% \\
 & Max Peak & 44.01 MB & 80.67 MB & 45.4\% \\
\midrule
\multirow{2}{*}{BIRD} & Avg Peak & 51.76 MB & 93.32 MB & 44.5\% \\
 & Max Peak & 94.31 MB & 145.13 MB & 35.0\% \\
\midrule
\multirow{2}{*}{Spider 2.0} & Avg Peak & 66.52 MB & 340.57 MB & 80.5\% \\
 & Max Peak & 284.44 MB & 1027.88 MB & 72.3\% \\
\bottomrule
\end{tabular}
}
\caption{GPU memory consumption profiling on the test sets. Avg Peak refers to the average maximum memory allocated per sample, and Max Peak denotes the single highest memory spike observed during inference.}
\label{tab:memory_profiling}
\end{table}

\section{GPU Memory Profiling and Efficiency Analysis}
\label{sec:memory_profiling}

Processing massive relational databases presents significant memory challenges because the aggregated input sequence length can easily exceed the strict context limits of standard encoders.
Direct processing of all tables incurs quadratic memory costs due to the self-attention mechanism, leading to potential instability.

To demonstrate the effectiveness of our design, we compared our sliding-window strategy against a baseline that processes all retrieved tables simultaneously without a sliding window.
As shown in Table~\ref{tab:memory_profiling}, the efficiency gain is most pronounced in the Spider 2.0 dataset, which contains the largest and most complex schemas.
Under this setting, ATR achieves an 80.5\% reduction in average peak memory and completely prevents the massive memory spikes exceeding 1GB that are observed in the baseline.
These findings confirm that the sliding-window strategy effectively bounds the peak memory per forward pass, ensuring stability and preventing out-of-memory errors even for extremely large-scale retrieval tasks.

\begin{algorithm}[t]
\caption{Adaptive Table Retrieval}
\label{alg:rerank}
\begin{algorithmic}[1]
\Require Query $q$, List of Table $\mathcal{C}$, Model $M$, Size of Window $W$, Size of Retention $R$, Number of Table $C$
\Ensure List of Ranked Table $\mathcal{C}'$

\State \textbf{Variables:}
\State \hspace{\algorithmicindent} $\mathcal{C}' \leftarrow \emptyset$, $\mathcal{C}_{retain} \leftarrow \emptyset$, $idx \leftarrow C$
\State \hspace{\algorithmicindent} $thr_{rank} \leftarrow 0$, $thr_{finalized} \leftarrow False$

\While{$idx > 0$}
    \If{$\mathcal{C}_{retain} = \emptyset$}
        \State $\mathcal{C}_{window} \leftarrow \mathcal{C}[idx-W:]$
        \State $idx \leftarrow idx - W$
    \Else
        % \State $\mathcal{C}_{window} \leftarrow \mathcal{C}[idx-(W-R):idx] + \mathcal{C}_{retain}$
        \State {\small $\mathcal{C}_{window} \leftarrow \mathcal{C}[idx-(W-R):idx] + \mathcal{C}_{retain}$}
        \State $idx \leftarrow idx - (W - R)$
    \EndIf

    \State $logit, score \leftarrow M(q, \mathcal{C}_{window})$
    \State $\mathcal{C}_{retain} \leftarrow Decend_{score}(\mathcal{C}_{window})[:R]$
    \If{\textbf{not} $thr_{finalized}$}
        \State $thr_{rank} \leftarrow Rank(logit) + idx$
        \If{$Rank(logit) > R$}
            \State $thr_{finalized} \leftarrow True$
        \EndIf
    \EndIf
    \State $\mathcal{C}' \leftarrow \mathcal{C}' + Ascend_{score}(\mathcal{C}_{window} \setminus \mathcal{C}_{retain})$
\EndWhile

\State $\mathcal{C}' \leftarrow \mathcal{C}' + Ascend_{score}(\mathcal{C}_{retain})$
\State $\mathcal{C}' \leftarrow Reverse(\mathcal{C}')[:thr_{rank}-1]$
\State \Return $\mathcal{C}'$

\end{algorithmic}
\end{algorithm}

\section{Case Study}
We investigate the limitations of top-$k$ approaches through qualitative analysis on specific examples from the BIRD and Spider 2.0 datasets.
In these examples, relevant tables are retrieved using Contriever ($k=5$) and ATR.
SQL queries are generated using Qwen2.5-Coder-32B-Instruct.

As illustrated in Table~\ref{tab: case study 1: k as noise}, the top-$k$ approach results in retrieving unnecessary tables when the number of required tables is fewer than $k$.
These irrelevant tables can be noise, leading to confusion and incorrect SQL generation.
Conversely, as illustrated in Table~\ref{tab: case study 2: small k}, when the query necessitates more tables than $k$, top-$k$ retrieval fails to retrieve all essential tables, again resulting in incorrect SQL outputs.
In contrast, ATR adaptively retrieves appropriate tables based on a query, effectively retrieving all necessary tables while minimizing irrelevant ones. 
This demonstrates ATR's clear advantage of providing precise and optimized input for the generator, significantly improving the accuracy and reliability of the generated SQL query. 

To further demonstrate the robustness of our dynamic thresholding mechanism, we present three additional edge cases drawn from complex enterprise scenarios. As summarized in Table~\ref{tab:edge_cases}, these include: (1) cross-database joins, where queries require combining tables across different underlying datasets; (2) ambiguous table references, where identical table names appear in multiple databases and demand context-aware disambiguation; and (3) large-scale retrieval scenarios, where the number of gold tables far exceeds typical $k$ values, causing fixed top-$k$ approaches to structurally fail.

% Requires in preamble:
% \usepackage{booktabs, xcolor, array}
\definecolor{atrgreen}{RGB}{0,128,0}
\definecolor{topkred}{RGB}{180,50,50}

\begin{table*}[t]
\centering
\small
\setlength{\tabcolsep}{4pt}
\renewcommand{\arraystretch}{1.2}
\begin{tabular}{@{} p{1.3cm} >{\raggedright\arraybackslash}p{13.8cm} @{}}
\toprule

\multicolumn{2}{@{}l}{\textbf{Case 1: Cross-Database Joins}} \\
\midrule
\textbf{Query} & \textit{``Could you provide, for the United States, France, China, Italy, Spain, Germany, and Iran, the total number of confirmed COVID-19 cases as of April 20, 2020, along with the number of cases per 100,000 people based on their total 2020 populations calculated by summing all relevant population entries from the World Bank data''} \\[3pt]
\textbf{Gold} & \texttt{covid19\_jhu\_csse.summary}, \texttt{world\_bank\_wdi.indicators\_data} \\[3pt]
\textbf{ATR} & \textcolor{atrgreen}{\texttt{covid19\_jhu\_csse.summary}}, \textcolor{atrgreen}{\texttt{world\_bank\_wdi.indicators\_data}} \\[3pt]
\textbf{Top-5} & \textcolor{topkred}{\texttt{world\_bank\_health\_population.health\_nutrition\_population}},\newline \textcolor{topkred}{\texttt{world\_bank\_health\_population.health\_nutrition\_population}},\newline \textcolor{topkred}{\texttt{world\_bank\_health\_population.health\_nutrition\_population}},\newline \textcolor{topkred}{\texttt{world\_bank\_global\_population.population\_by\_country}},\newline \textcolor{topkred}{\texttt{world\_bank\_wdi.series\_times}} \\

\midrule
\multicolumn{2}{@{}l}{\textbf{Case 2: Ambiguous Table References}} \\
\midrule
\textbf{Query} & \textit{``Calculate the average sales per order for each customer within distinct RFM segments, considering only `delivered' orders''} \\[3pt]
\textbf{Gold} & \texttt{E\_commerce.orders}, \texttt{E\_commerce.order\_items}, \texttt{E\_commerce.customers} \\[3pt]
\textbf{ATR} & \textcolor{atrgreen}{\texttt{orders}} (E\_commerce), \textcolor{atrgreen}{\texttt{order\_items}} (E\_commerce), \textcolor{atrgreen}{\texttt{customers}} (E\_commerce) \\[3pt]
\textbf{Top-5} & \textcolor{topkred}{\texttt{orders}} (electronic\_sales), \textcolor{atrgreen}{\texttt{orders}} (E\_commerce), \textcolor{topkred}{\texttt{leads\_closed}} (E\_commerce), \textcolor{topkred}{\texttt{order\_payments}} (electronic\_sales), \textcolor{topkred}{\texttt{orders}} (delivery\_center) \\

\midrule
\multicolumn{2}{@{}l}{\textbf{Case 3: Large-Scale Retrieval}} \\
\midrule
\textbf{Query} & \textit{``Could you tell me the average number of engaged sessions per user for December 2020, counting only those sessions where the event parameter `session\_engaged' is equal to `1' and using `user\_pseudo\_id' combined with the `ga\_session\_id' to identify distinct sessions?''} \\[3pt]
\textbf{Gold} & \texttt{ga4.events\_20201201} \ldots\ \texttt{ga4.events\_20201231} (31 daily tables) \\[3pt]
\textbf{ATR} & \textcolor{atrgreen}{\texttt{events\_20201201}}, \textcolor{atrgreen}{\texttt{events\_20201202}}, \ldots, \textcolor{atrgreen}{\texttt{events\_20201231}} (31/31 retrieved) \\[3pt]
\textbf{Top-20} & \textcolor{atrgreen}{\texttt{events\_20201201}}, \ldots, \textcolor{atrgreen}{\texttt{events\_20201211}} (11/31 retrieved) \\

\bottomrule
\end{tabular}
\caption{Edge-case retrieval results. \textcolor{atrgreen}{Green} denotes correctly retrieved gold tables; \textcolor{topkred}{red} denotes irrelevant or missed tables.}
\label{tab:edge_cases}
\end{table*}

% Although ATR demonstrates substantial improvements in both retrieval and downstream execution accuracy, two limitations remain. 
% First, the sliding window re-ranking method effectively reduces computational complexity, but determining optimal window sizes and retention parameters may necessitate additional empirical tuning across different datasets or retrieval scenarios. 
% Second, ATR currently targets structured tabular data exclusively, and extending its adaptive retrieval strategy to handle other data modalities or mixed data types remains an open research challenge. 
% Nevertheless, by establishing a robust framework for adaptively retrieving relevant information efficiently without direct interaction with LLMs, our methodology provides a strong foundation for future studies aiming to generalize retrieval-augmented generation across diverse data types and broader application contexts. 

% \begin{figure*}[ht]
%     \centering
%     \includegraphics[width=0.55\textwidth]
%     {latex/images/supplymentary_sliding_window.pdf}
%     \caption{Example for re-ranking tables. ATR refines the initial ranking using its sliding window re-ranking algorithm. Since re-ranking is performed in overlapping window-sized segments rather than processing all tables simultaneously, ATR avoids issues related to input length constraints. Thus, there is no limitation on the number of input tables in our approach}
%     \label{fig:Example for re-ranking tables}
% \end{figure*}

\begin{table*}[ht]
\centering
% \toprule
% \hline
\begin{tabular}{p{0.48\textwidth}p{0.48\textwidth}}
% \hline
\toprule
\multicolumn{2}{p{0.96\textwidth}}{
\textbf{Question}\vspace{2pt}} \\
% \hline
% \toprule
\bottomrule
\multicolumn{2}{p{0.96\textwidth}}{
\begin{minipage}[c]{0.96\textwidth}
\vspace{6pt}
Please list player names which are higher than 180.
\vspace{6pt}
\end{minipage}
} 
% \vspace{0.01\textheight}
\\

% \hline
\toprule
\textbf{Tables Retrieved by ATR} & \textbf{Tables Retrieved by Top-$k$} \\
% \hline
\bottomrule

\begin{minipage}[t]{0.48\textwidth}
"Player"
% ["european\_football\_2", 
% "Player"]
\end{minipage} 
&
\begin{minipage}[c]{0.48\textwidth}
\vspace{6pt}
"Player" \newline
\textcolor{red}{"Match"} \newline \textcolor{red}{"League"} \newline \textcolor{red}{"Team"} \newline  \textcolor{red}{"superhero"}
% \includegraphics[width=0.86\textwidth]
% {latex/images/case_study_ours_1.pdf}
% ["european\_football\_2", "Player"] \newline 
\vspace{6pt}
\end{minipage} \\
% \hline
\toprule
\textbf{SQL Generated by ATR} & \textbf{SQL Generated by Top-$k$} \\
% \hline
\bottomrule
\begin{minipage}[c]{0.48\textwidth}
%SELECT player\_name FROM Player WHERE height > 180 
\vspace{6pt}
\includegraphics[width=0.86\textwidth]
{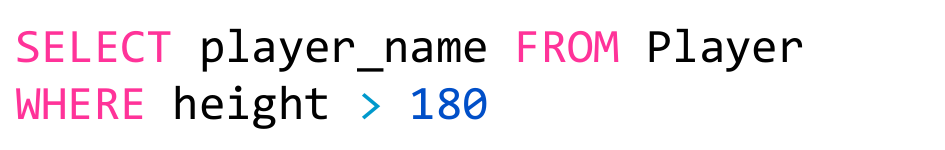}
% "Player" \newline 
% \textcolor{red}{"Match"} \textcolor{red}{"Team"} 
% \newline
% \textcolor{red}{"League"} \textcolor{red}{"superhero"}
\vspace{6pt}
\end{minipage}
&
\begin{minipage}[c]{0.48\textwidth}
%SELECT T1.player\_name FROM Player AS T1 INNER JOIN Match
%AS T2 ON T1.player\_api\_id = T2.home\_player\_1 WHERE T1.height > %180 
\includegraphics[width=1.5\textwidth]
{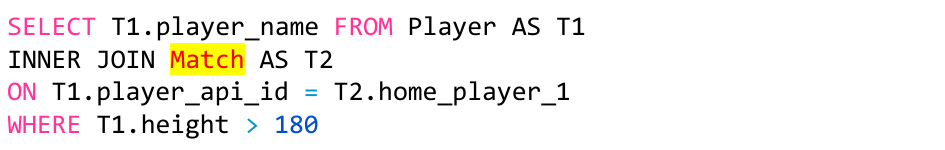}
\end{minipage} \\
% \hline
\toprule
\multicolumn{2}{p{0.96\textwidth}}{\textbf{Retrieved Table Schema}} \\
% \hline
\bottomrule
\multicolumn{2}{p{0.96\textwidth}}{
\begin{minipage}[c]{0.96\textwidth}
\vspace{6pt}
"Player": \textit{id}, \textit{player\_api\_id}, \textit{height}, \textit{weight}, \textit{birthday}, \textit{player\_name}, \textit{player\_fifa\_api\_id} \newline
\textcolor{red}{"Match": \textit{id}, \textit{home\_player\_1}, \textit{stage}, \textit{goal}, \textit{season}, \textit{country\_id}, \textit{league\_id}, \textit{match\_api\_id}, ...} \newline
\textcolor{red}{"League": \textit{id}, \textit{country\_id}, \textit{name} } \newline
\textcolor{red}{"Team": \textit{id}, \textit{team\_api\_id}, \textit{team\_fifa\_api\_id}, \textit{team\_long\_name}, \textit{team\_short\_name}}  \newline
\textcolor{red}{"superhero": \textit{id}, \textit{superhero\_name}, \textit{full\_name}, \textit{gender\_id}, \textit{eye\_colour\_id}, \textit{hair\_colour\_id}, ...}
% ["european\_football\_2", "Player"]|: Player.id, Player.player\_api\_id, Player.height .. \newline
% ["european\_football\_2", "Match"]: Match.home\_player\_1, Match.match\_api\_id, Match.goal, .. \newline
% ["european\_football\_2", "Team"]: Team.id, Team.team\_api\_id, ..  \newline
% ["european\_football\_2", "League"]: League.id, League.country\_id, ..  \newline
% ["superhero", "superhero"]: superhero.id, superhero.superhero\_name, superhero.height\_kg, ..
\vspace{6pt}
\end{minipage}
} \\
% \hline
\bottomrule
\end{tabular}
\caption{
A case study illustrating a scenario where the retrieval size $k$ exceeds the number of tables required to resolve the query. As the top-$k$ approach retrieves unnecessary tables, it propagates noise, degrading downstream SQL generation. Irrelevant tables are in red, and unnecessary tables that do not contribute to answering the query are highlighted in yellow. Retrieved table schema displays schema information for all retrieved tables.
% irrelevant한 table로 인해 downstream task degradation 하는 케이스
% Scenario where fixed $k$ exceeds the required number of tables. Tables retrieved by the top-$k
% $ method but irrelevant to SQL generation are indicated in red text. Text highlighted in yellow represents unnecessary tables that do not contribute to answering the question. Retrieved Table Schema displays schema information for all retrieved tables.
% Retrieved Table Schema denotes the schema information for tables retrieved by ATR and top-$k$.
} 
\label{tab: case study 1: k as noise}
\end{table*}
% \begin{figure*}[ht]
%     \centering
%     \includegraphics[width=0.9\textwidth]{latex/images/case_study_top_k_exceeds.pdf}
%     \caption{A case study illustrating a scenario where the retrieval size $k$ exceeds the number of tables required to resolve the query. As the top-$k$ approach retrieves unnecessary tables, it propagates noise, degrading downstream SQL generation. Irrelevant tables are in red, and unnecessary tables that do not contribute to answering the query are highlighted in yellow. Retrieved table schema displays schema information for all retrieved tables.}
%     \label{fig: case study 1: k as noise}
% \end{figure*}
\begin{table*}[ht]
\centering
\begin{tabular}{p{0.48\textwidth}p{0.48\textwidth}}
% \hline
\toprule
\multicolumn{2}{p{0.96\textwidth}}{\textbf{Question}} \\
\bottomrule
% \hline
\multicolumn{2}{p{0.96\textwidth}}{
\begin{minipage}[c]{0.96\textwidth}
\vspace{6pt}
Please calculate the monthly average levels of PM10, PM2.5 FRM, PM2.5 non-FRM, volatile organic emissions, SO2 (scaled by a factor of 10), and Lead (scaled by a factor of 100) air pollutants in California for the year 2020.
\vspace{6pt}
\end{minipage}} \\
% \hline
\toprule
\textbf{Tables Retrieved by ATR} & \textbf{Tables Retrieved by Top-$k$} \\
% \hline
\bottomrule
\begin{minipage}[c]{0.48\textwidth}
\vspace{6pt}
"pm25\_frm\_daily\_summary" \newline
"lead\_daily\_summary" \newline
"pm10\_daily\_summary" \newline
"pm25\_nonfrm\_daily\_summary" \newline
"so2\_daily\_summary" \newline
"voc\_daily\_summary"
\vspace{6pt}
\end{minipage} 
&
\begin{minipage}[c]{0.48\textwidth}
\vspace{6pt}
"pm25\_frm\_daily\_summary" \newline
"lead\_daily\_summary" \newline
"pm10\_daily\_summary" \newline
\textcolor{red}{"pm25\_frm\_hourly\_summary"} \newline
\textcolor{red}{"pm10\_hourly\_summary"} \newline
\textcolor{lightgray}{"pm25\_nonfrm\_daily\_summary"} \newline
\textcolor{lightgray}{"so2\_daily\_summary"} \newline
\textcolor{lightgray}{"voc\_daily\_summary"}
\vspace{6pt}
\end{minipage} \\
% \hline
\toprule
\textbf{SQL Generated by ATR} & \textbf{SQL Generated by Top-$k$} \\
% \hline
\bottomrule
%SELECT player\_name FROM Player WHERE height > 180 
\vspace{1pt}
\includegraphics[width=0.48\textwidth]
{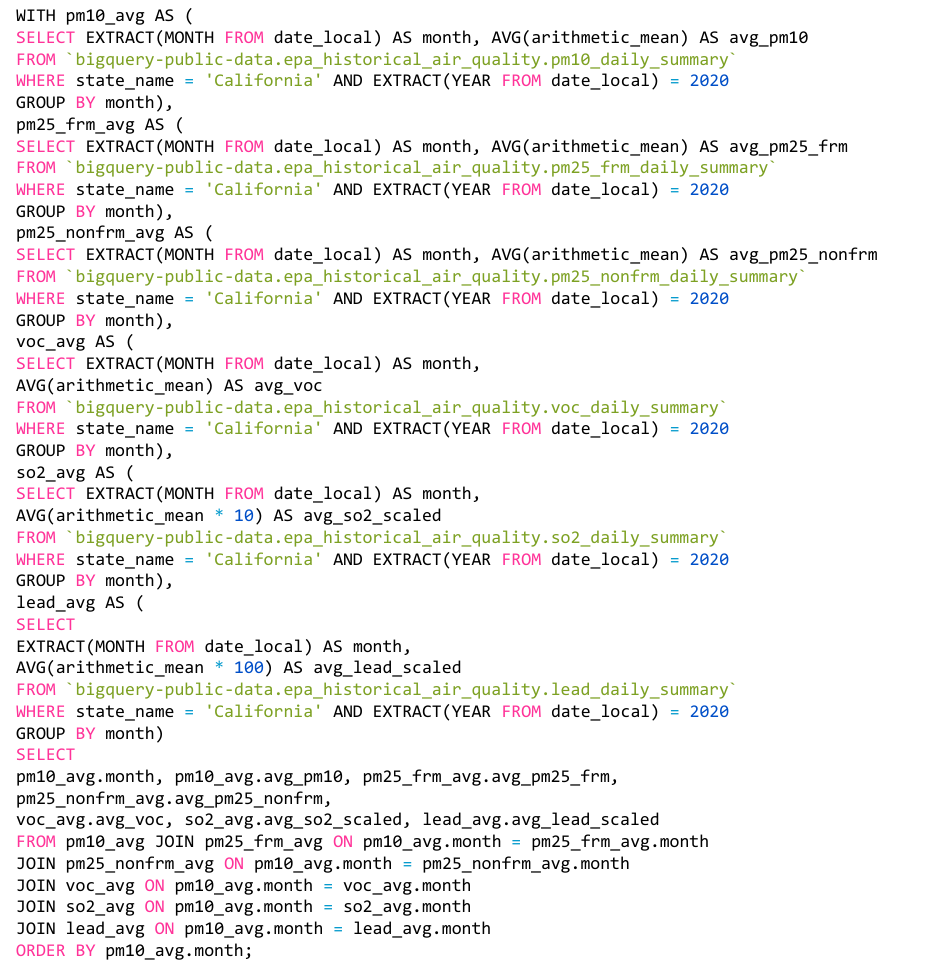}
\vspace{1pt}
&
%SELECT T1.player\_name FROM Player AS T1 INNER JOIN Match
%AS T2 ON T1.player\_api\_id = T2.home\_player\_1 WHERE T1.height > %180 
\vspace{1pt}
\includegraphics[width=0.48\textwidth]
{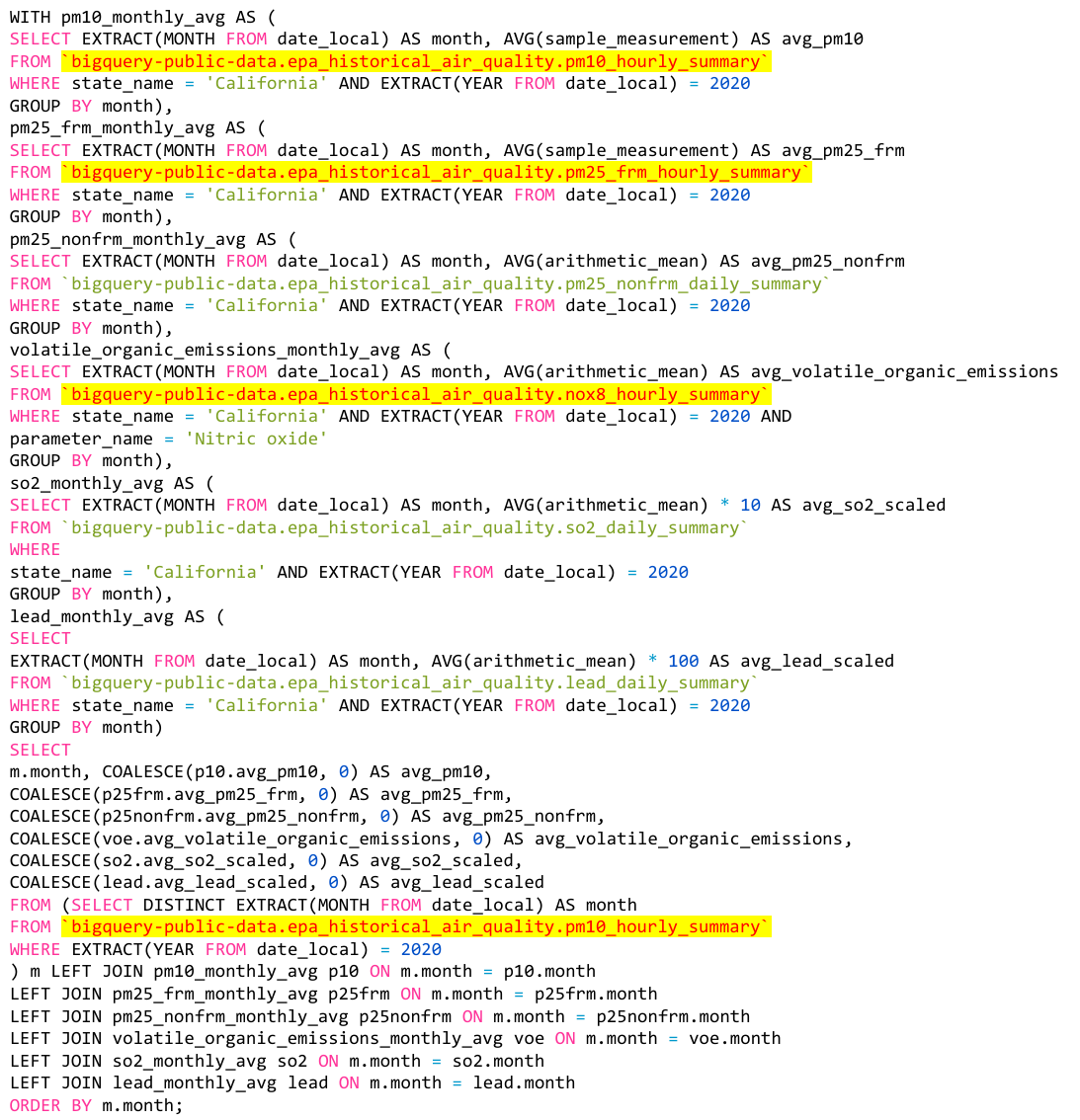}
\vspace{1pt}\\
% \hline
\toprule
\multicolumn{2}{p{0.96\textwidth}}{\textbf{Retrieved Table Schema}} \\
% \hline
\bottomrule
\multicolumn{2}{p{0.96\textwidth}}{
\begin{minipage}[c]{0.96\textwidth}
\vspace{2pt}
"pm25\_frm\_daily\_summary": \textit{arithmetic\_mean}, \textit{date\_local}, \textit{state\_code}, \textit{county\_code}, ... \newline
"lead\_daily\_summary": \textit{arithmetic\_mean}, \textit{date\_local}, \textit{state\_code}, \textit{county\_code}, ...\newline
"pm10\_daily\_summary": \textit{arithmetic\_mean}, \textit{date\_local}, \textit{state\_code}, \textit{county\_code}, ... \newline
"pm25\_nonfrm\_daily\_summary": \textit{arithmetic\_mean}, \textit{date\_local}, \textit{state\_code}, \textit{county\_code}, ... \newline
"so2\_daily\_summary": \textit{arithmetic\_mean}, \textit{date\_local}, \textit{state\_code}, \textit{county\_code}, ... \newline
"voc\_daily\_summary": \textit{arithmetic\_mean}, \textit{date\_local}, \textit{state\_code}, \textit{county\_code}, ... \newline
\textcolor{red}{"pm25\_frm\_hourly\_summary": \textit{date\_local}, \textit{sample\_measurement}, \textit{state\_code}, \textit{county\_code},  ...} \newline
\textcolor{red}{"pm10\_hourly\_summary": \textit{date\_local}, \textit{sample\_measurement}, \textit{state\_code}, \textit{county\_code}, ...}
\vspace{2pt}
\end{minipage}
} \\
% \hline
\bottomrule
\end{tabular}
\caption{
A case study illustrating a scenario where the retrieval size $k$ is smaller than the number of tables required to resolve the query. As the top-
$k$ approach fails to retrieve all necessary tables, it produces inaccurate SQL, whereas ATR retrieves all essential tables, enabling correct SQL generation. Tables not retrieved but crucial for SQL generation are displayed in light gray.
% 필요한 table을 뽑지못해 downstream task를 수행하지 못하는 상황
% Scenario where more tables are needed than the fixed $k$. Neccessary but not retrieved by the top-$k$ method are not marked in lightgray text.
% Unnecessary tables retrieved by the top-$k$ method are marked in red text. Text highlighted in yellow in top-$k$ SQL indicates tables unnecessary for answering the question. 
% \textbf{Retrieved Table Schema} denotes the schema information for tables retrieved by ATR and top-$k$.
}
\label{tab: case study 2: small k}
\end{table*}

\begin{figure*}[ht]
    \centering
    \includegraphics[width=0.65\textwidth]{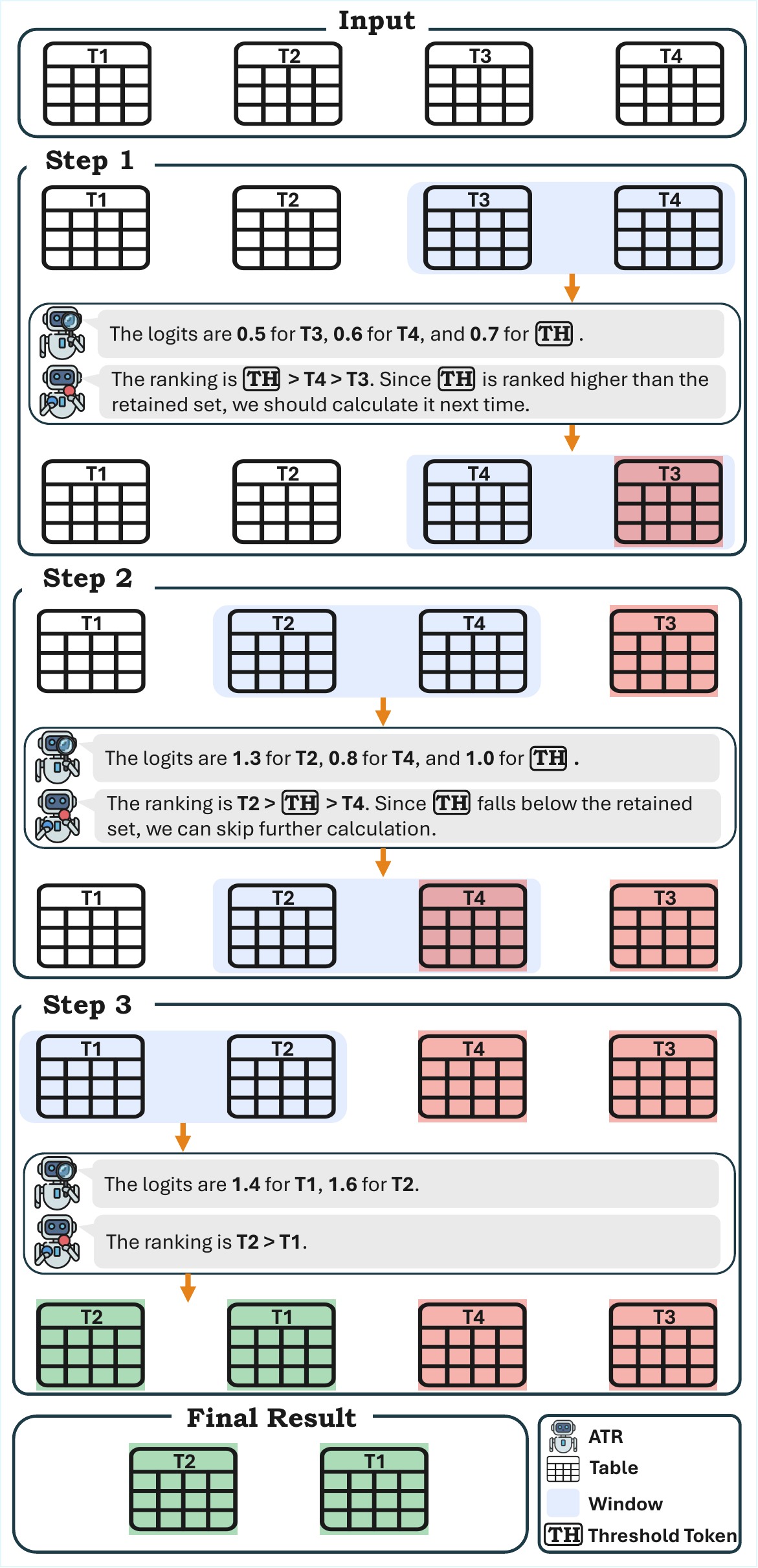}
    \caption{An illustrative example of the sliding window reranking process in ATR with four input tables, window size $W=2$, and retention size $R=1$. Tables highlighted in red indicate those whose rankings are finalized but excluded from the final set, while tables in green represent the tables that are ultimately retrieved. The input tables are initially ordered by query-table similarity in descending order from left to right. ATR processes tables starting from the lowest-ranked ones (rightmost) within the window. In each step, ATR compares the logits of table tokens with the threshold token ($T_{th}$) and retains the top-$R$ tables for the next iteration. }
    \label{fig:sliding_window_example}
\end{figure*}

\begin{figure*}[ht]
    \centering
    \includegraphics[width=\textwidth]{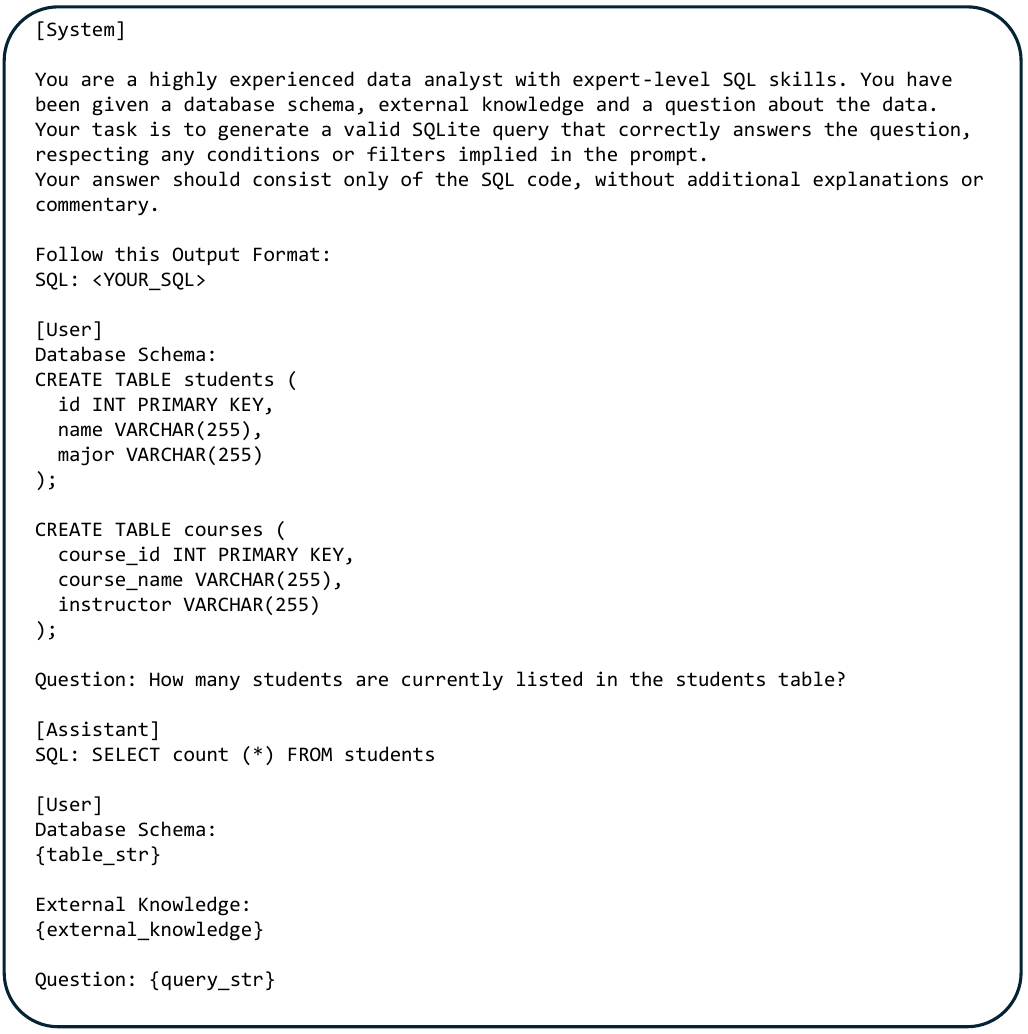}
    \caption{Prompt template for Spider and BIRD datasets.}
    \label{fig:Prompt for Spider and BIRD with external knowledge}
\end{figure*}

\begin{figure*}[ht]
    \centering
    \includegraphics[width=\textwidth]{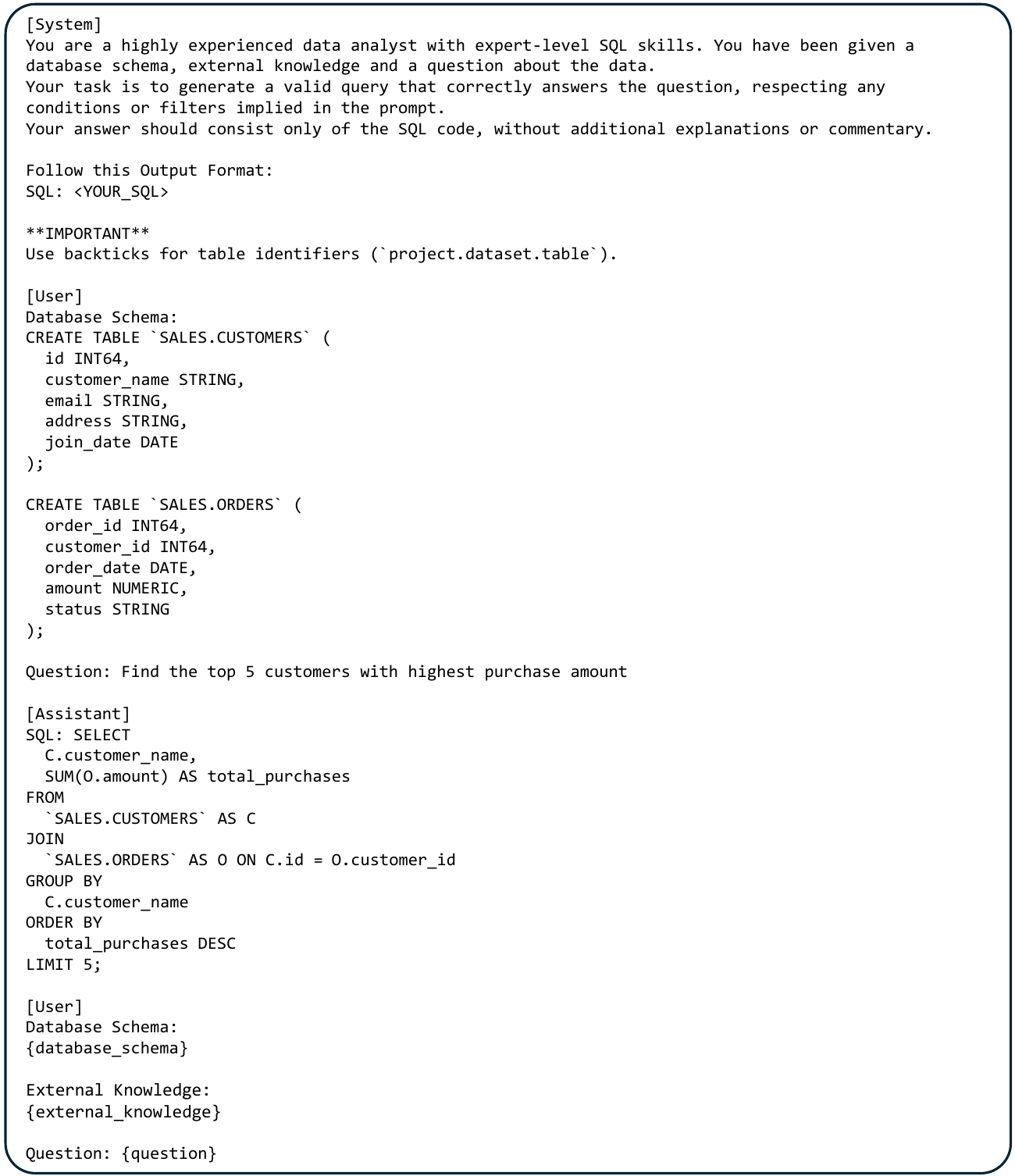}
    \caption{Prompt template for Spider 2.0 (BigQuery dialect)}
    \label{fig:BigQuery Prompt Examples}
\end{figure*}

\begin{figure*}[ht]
    \centering
    \includegraphics[width=0.9\textwidth]{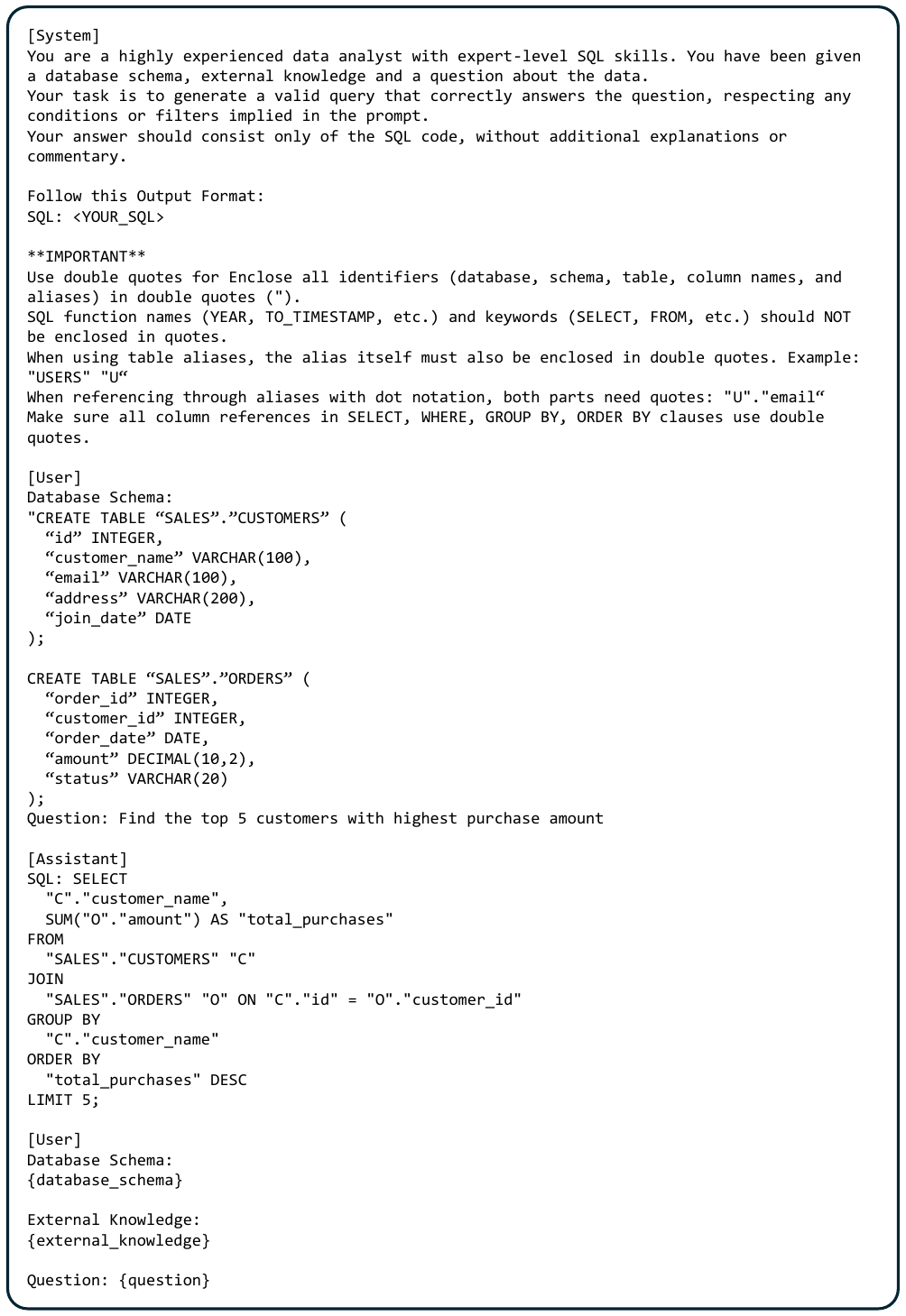}
    \caption{Prompt template for Spider 2.0 (Snowflake dialect)}
    \label{fig:Snowflake Prompt Examples}
\end{figure*}

\begin{figure*}[ht]
    \centering
    \includegraphics[width=\textwidth]{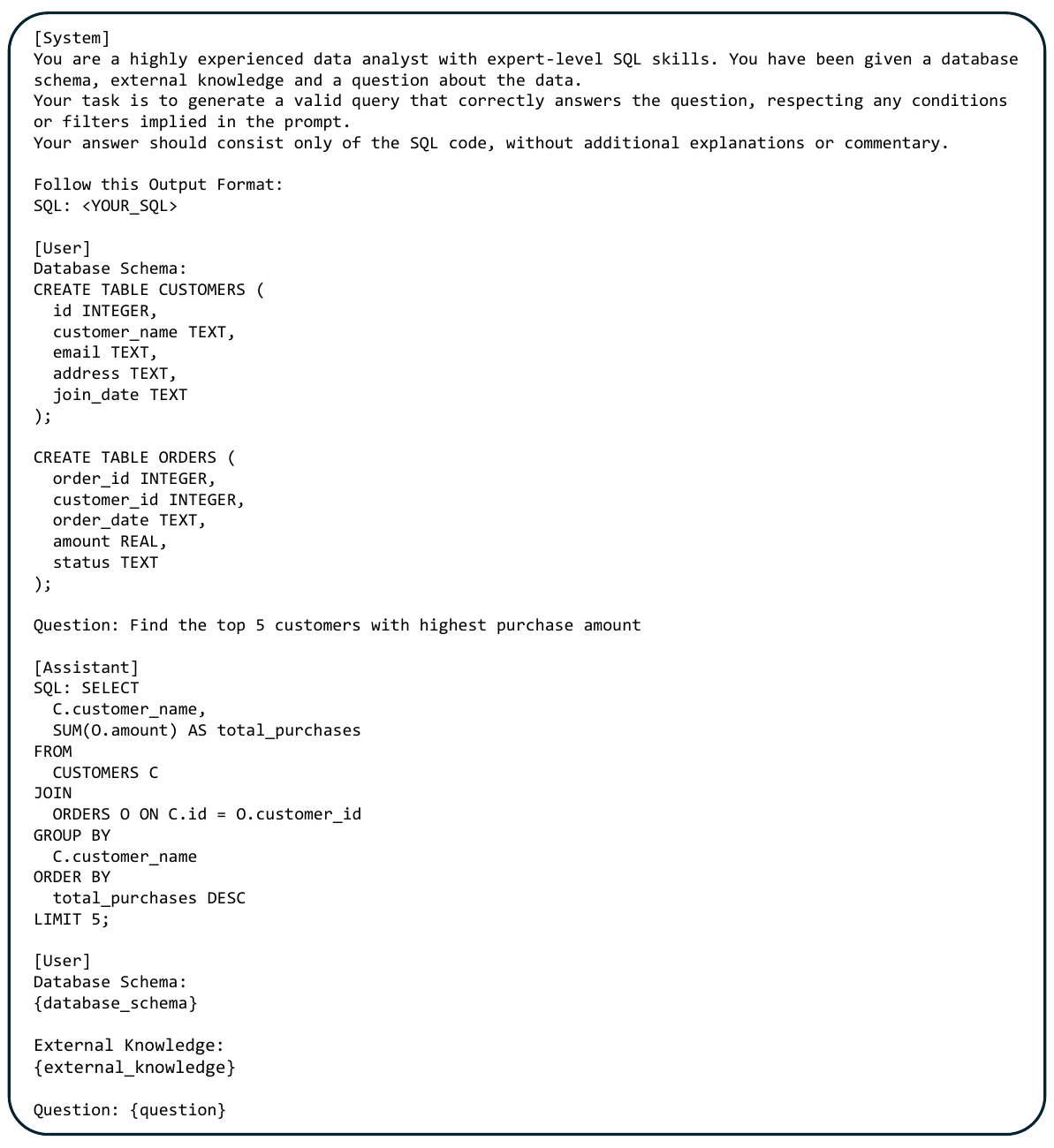}
    \caption{Prompt template for Spider 2.0 (SQLite dialect)}
    \label{fig:SQLite Prompt Examples}
\end{figure*}

\end{document}